\documentclass[a4paper,11pt]{article}
\pdfoutput=1 

\usepackage{jheppub} 

\usepackage[T1]{fontenc} 

\usepackage{tabularx}
\usepackage{graphicx}
\usepackage{epsf,color}
\usepackage[dvipsnames]{xcolor}
\usepackage{caption}
\usepackage{subcaption}
\usepackage{lmodern}
\usepackage[stretch=10,shrink=10]{microtype}
\definecolor{nicered}{rgb}{0.5,0.,0.}
\definecolor{nicegreen}{rgb}{0.,0.5,0.}
\definecolor{niceblue}{rgb}{0.,0.,0.5}
\hypersetup{colorlinks,citecolor=nicegreen,linkcolor=nicered,urlcolor=niceblue}

\newcommand{\GeV}{\mathrm{GeV}}
\newcommand{\TeV}{\mathrm{TeV}}

\newcommand{\mumut}{\ensuremath{\mu^+\mu^-\to}}

\newcommand{\orcidPS}{\href{https://orcid.org/0000-0002-5976-0317}{\includegraphics[height=9pt]{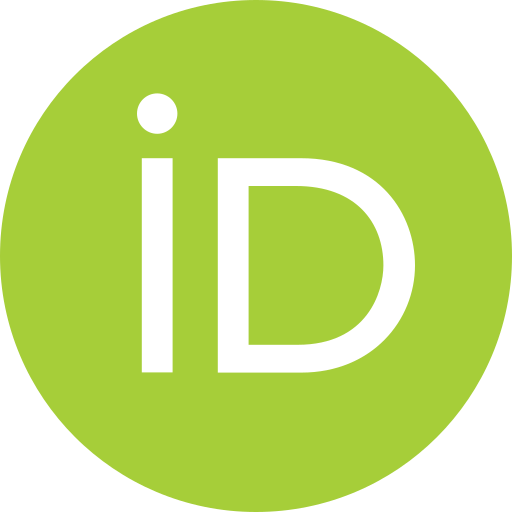}~}}
\newcommand{\orcidJR}{\href{https://orcid.org/0000-0003-1866-0157}{\includegraphics[height=9pt]{orcid.png}~}}
\newcommand{\orcidPB}{\href{https://orcid.org/0000-0003-4579-0387}{\includegraphics[height=9pt]{orcid.png}~}}

\title{NLO Electroweak Corrections to Multi-Boson Processes at a Muon Collider}

\author[a,1]{Pia M. Bredt\orcidPB,\note{Corresponding author.}}
\author[b]{Wolfgang Kilian,}
\author[a]{J\"urgen Reuter\orcidJR}
\author[a]{and Pascal Stienemeier\orcidPS}

\affiliation[a]{Deutsches Elektronen-Synchrotron DESY, Notkestr.~85,
  22607 Hamburg, Germany}
\affiliation[b]{Department of Physics, University of Siegen,
  Walter-Flex-Str.~3, 57068 Siegen, Germany}

\emailAdd{pia.bredt@desy.de}
\emailAdd{kilian@physik.uni-siegen.de}
\emailAdd{juergen.reuter@desy.de}
\emailAdd{pascal.stienemeier@desy.de}

\preprint{DESY 22-124, SI-HEP-2022-24}

\abstract{We present results on NLO electroweak (EW) corrections
  to multiple massive boson production processes at a future muon
  collider. Inclusive cross sections with $\mathcal{O}(\alpha)$
  corrections for processes for up to four bosons in the final
  state as well as differential distributions for $HZ$ production are
  computed for $\sqrt{s}=3$, $10$ and $14$ TeV by using FKS
  subtraction in the NLO EW automated Monte-Carlo framework
  \texttt{WHIZARD+RECOLA}. Large logarithmic effects due to collinear
  ISR and EW virtual correction factors as well as the impacts of an
  energy cut on hard photons are discussed with an emphasis on the
  properties of Higgsstrahlung. The potential of a proposed muon
  collider for studying physics of the EW sector is underlined by the
  EW corrections significantly affecting observables for processes at
  high energies and boson multiplicities.}

\begin{document}
\maketitle
\flushbottom

\section{Introduction}

Our current understanding of electroweak (EW) physics is governed by the
Standard Model (SM).  The model predicts a simple symmetry structure
at the fundamental level, and it yields the correct low-energy limit
of electromagnetic, weak, and strong interactions.
Yet, genuine EW
and Higgs data can justify this theoretical concept only on a
rather superficial level.  Energy-frontier experiments, in particular
at the Large Hadron Collider (LHC), do not access energy scales in
EW interactions beyond about $1\,\TeV$.  This is less than
one order of magnitude above the inherent mass scale of
$v=(\sqrt2 G_F)^{-1/2}=246\;\GeV$.  The highest multiplicity of the
mutual interactions of massive EW particles ($t,W,Z,H$) that
can be probed is three or, in a few cases, four.

If the SM and its symmetry structure are to be established on a deeper
level, or modified by new effects at short distance, there is a clear
need for collider experiments which are able to directly probe
EW interactions at higher energy and higher multiplicity, and
simultaneously with better precision than possible today.  There
are plans for a next-generation hadron collider~\cite{Abada:2019lih,
Benedikt:2018csr}.  Modern
lepton-acceleration technology (ILC~\cite{Baer:2013cma,Behnke:2013lya},
CLIC~\cite{CLIC:2016zwp,Aicheler:2012bya}) would also allow us to further
extend the energy frontier.  Recently, colliders based on storage
rings with muons have been proposed as a further option for exploring
the multi-$\TeV$ range, exploiting new ideas for the cooling of the
high-enery, high-intensity muon beams~\cite{Delahaye:2019omf,
Bartosik:2020xwr,Schulte:2020xvf,Long:2021upy,Aime:2022flm,
MuonCollider:2022xlm}.

A muon collider would indeed become a valuable tool for pushing the
limits of the SM: center-of-mass (c.m.) energies beyond $10\;\TeV$ are
considered technologically feasible.  Proposed benchmark values which
we will use in this work, are $3\;\TeV$, $10\;\TeV$, and $14\;\TeV$.
The clean lepton-collider environment in conjunction with dedicated
detectors will enable exclusive measurements of final states in both
leptonic and hadronic channels.

Muons are elementary particles in the SM, so their collisions offer
remarkable options for the search for new particles and interactions
at the full machine energy.  The physics accessible at a muon collider
is very similar to the physics at an $e^+e^-$ collider with the same
energy (e.g., CLIC with $\sqrt{s}=3\;\TeV$)~\cite{MuonCollider:2022xlm}.
Moreover, the larger mass
of the muon as compared to the electron, reduces dominant effects
that dilute leptonic collisions, namely beamstrahlung as a collective
beam-beam interaction, and initial-state radiation (bremsstrahlung) as
a reduction of the effective c.m.\ energy due to photon emission.

A high-energy muon collider will thus open up possibilities of
studying a rich set of processes with single and multiple EW
gauge and Higgs bosons.  In this work, we focus on multi-boson
production processes $\mu^+\mu^- \to V^n H^m$ with $V \in \{W^\pm,Z\}$
and $n+m \leq 4$.  These processes allow us to scrutinize
the EW gauge and symmetry-breaking sector, study Higgs interactions in
detail, and search for new (heavy or light) states which couple to the
EW sector.

At leading order in the SM couplings, standard universal Monte-Carlo
(MC) programs provide detailed predictions for this class of
processes, both within the SM and in perturbative extensions such as
the Standard-Model Effective Field Theory (SMEFT).  In
Ref.~\cite{CLIC:2018fvx}, we computed multi-boson production cross
sections with $n+m\leq 4$ for the CLIC $e^+e^-$ collider with
$\sqrt{s}\leq 3\;\TeV$.  The results also apply to a muon collider in
the same energy range.  A dedicated muon-collider
study~\cite{Costantini:2020stv} furthermore covered energies up to
$30\;\TeV$.  The specific phenomenology of lepton spin-flip
interactions which distinguishes muons from electrons, was studied for
multi-boson production processes in Refs.\
\cite{Han:2020pif,Chiesa:2020awd,Dermisek:2021mhi,Han:2021lnp}.

In order to establish a potential deviation from the predictions of
the SM and to claim a potential discovery of new
physics, predictions of the SM (or any other reference model) have to
be provided with a precision that is at least as good as the combined
statistical and systematic uncertainty of the experimental
measurement. Due to the high envisaged integrated luminosities and the
immense precision of modern highly-granular particle physics detectors
with analyses based on particle flow, leading-order (LO) calculations
are almost always insufficient to match the experimental precision on
the theory side.  To this end, in this paper we complement the
previous results by cross-section results and exemplary distributions
at next-to-leading order (NLO) in the complete SM.

For asymptotically high energy, NLO corrections to exclusive final
states are dominated by Sudakov-type double logarithms.  The
correction to the inclusive EW-singlet total cross section depends
only on the SM running couplings as functions of energy, with a rather
moderate energy dependence.  However, the
multiplicity distribution of the exclusive final states within this
total rate exhibits a transition to jet-like EW
radiation patterns as soon as the vector-boson masses become
negligible in the multi-$\TeV$ regime.  By unitarity, the appearance
of high-multiplicity final states is compensated by an eventual
logarithmic reduction ($\log s/m_V^2$) of low-multiplicity final
states.  This reduction can be exponentiated to leading-logarithmic
order by standard techniques.  A simple estimate suggests EW-jet
dominance and strong Sudakov suppression at the highest muon-collider
energies (e.g., $30\;\TeV$).  Conversely, at $3\;\TeV$ higher-order
Sudakov logarithms are still subleading compared to standard NLO
perturbative corrections.  In this work, we compute the complete
fixed-order NLO corrections to multi-boson production processes for a
muon collider in the transition region $3\dots 14\;\TeV$, where both
logarithmic and non-logarithmic NLO contributions are important.

After a brief overview in Sec.~\ref{secSetup} on our setup and methods
used for the NLO EW computations we present results on inclusive cross
sections in Sec.~\ref{secinclusiveresults}. Included are cross section
scans at NLO EW in $\sqrt{s}$ for $HZ$ and $ZZ$ production and more
general results to two, three and four bosons in the final state at 3
TeV, and additionally to two and three bosons at 10 and 14 TeV muon
collider, respectively. Moreover, in this section an estimate for
the ISR effects on the corresponding processes is given by LO cross
section results including leading logarithmic (LL) resummation in
$\alpha$ of the initial state.  In Sec.~\ref{differentialresults}
differential distributions for the process $\mu^+\mu^-\rightarrow HZ$
at the three proposed muon collider energies with and without cuts on
hard photons occuring at NLO in $\alpha$ are discussed.

\section{Setup and NLO framework}
\label{secSetup}

We compute NLO cross sections and distributions for multi-boson
processes using the Monte-Carlo event generator
\texttt{WHIZARD}~\cite{Kilian:2007gr,Moretti:2001zz}.  This generator
is a multi-purpose program for cross-section and distribution
calculation as well as for generating simulated event samples; for a
recent application to muon-collider EW and Higgs physics cf.\
\cite{Han:2021lnp}.  We recently extended \texttt{WHIZARD}'s automated
framework to account for the complete perturbative NLO corrections in
the full SM.  While the new NLO module will be documented in detail in
a separate publication~\cite{WhizardNLO}, we summarize
methods and features below.

The EW one-loop virtual contributions are provided by \texttt{RECOLA}
\cite{Actis:2016mpe}, which can account for the full mass dependence
of fermions and bosons.  Additionally, for NLO QED cross sections for
$HZ$ and $ZZ$, we have used our interface to the one-loop provider
\texttt{OpenLoops} \cite{Buccioni:2019sur}.  Phase-space construction and
subtraction follow the FKS
scheme~\cite{Frixione:1995ms,Frixione:1997np} (for NLO QCD in the color-flow
formalism~\cite{Kilian:2012pz}), fully automatized for lepton and
hadron colliders~\cite{ChokoufeNejad:2016qux,WhizardNLO}.\footnote{%
  Earlier versions of \texttt{WHIZARD} used special-tailored NLO-EW
  amplitudes~\cite{Kilian:2006cj, Robens:2008sa} or a different
  subtraction scheme for QCD-NLO
  corrections~\cite{Binoth:2009rv,Greiner:2011mp,Bach:2017ggt}.}  For
an appropriate FKS phase-space construction with massive particles in
both initial and final state, we adjust the mapping between the Born
and real-radiation phase-space parameterizations according to the
on-shell projection proposed in
\cite{Dittmaier:2015bfe,Denner:2000bj}.  We have generalized this
phase-space construction scheme from its original application to
factorized processes with massive resonances~\cite{Bach:2017ggt}.  The
integration proceeds via numerical phase-space sampling with
multi-channel adaption~\cite{Ohl:1998jn}.  The
time-consuming NLO calculations and simulations are heavily
facilitated using \texttt{WHIZARD}s MPI-based
parallelization~\cite{Brass:2018xbv}.

For the numerical results of this study, we define the electromagnetic
coupling $\alpha$ at the hard scale of the process in the $G_{\mu}$
input-parameter scheme, thus resumming a certain class of
logarithmically enhanced QED corrections.  Regarding massive vector
bosons, we impose on-shell renormalization conditions and set particle
widths to zero, thus maintaining EW gauge invariance in the
interference of $s$-channel and $t$-channel contributions.  Throughout
the calculation, we use nonzero masses for all particles except for
photon and neutrinos, and the corresponding Yukawa couplings are
included.  While for light quarks and electrons in loops this is
merely a technical detail without phenomenological significance,
keeping the muon mass nonzero regulates infrared and collinear
divergences associated with initial-state radiation.  In fact,
QED corrections beyond NLO are parameterically of order
$(\alpha/\pi)^2\log^2(s/m_\mu^2) \sim 0.1\;\%$ which is sufficiently
small in the present context.  This allows us to treat the colliding
$\mu^+\mu^-$ system perturbatively without the need for higher-order
resummation or for introducing leptonic parton distribution functions.

We performed technical sanity checks on the implemented FKS
subtraction scheme such as checking soft limits (note that for massive
emitters, there are no collinear subtractions) and cross checks
comparing the FKS real phase-space paramterization to the underlying
Born process $\mu^+\mu^- \to X$ with the LO parameterization of
$\mu^+\mu^- \to X + \gamma$ with a well-defined photon.

Beyond technical checks, we validated explicit NLO EW cross section
results of $e^+e^- \rightarrow HZ$ with
\texttt{MCSANCee}~\cite{Sadykov:2020dgm}. 
The electrons for these checks are treated as massive which is
analogous to the setup for processes with massive initial-state
muons.
The results and details of these rather technical checks
are deferred to the
appendix~\ref{validation}.

For reference, we list the numerical input parameters. They are used
consistently for LO and NLO amplitude calculations and for phase-space
construction, where applicable.
\begin{equation*}
  G_{\mu}= 1.166379\cdot 10^{-5} \; \text{GeV}^{-2} \\
\end{equation*}
\begin{eqnarray*}
  m_u &=\; \phantom{17}0.062           \;\text{GeV}\qquad\qquad
  m_d &=\; 0.083           \;\text{GeV} \\
  m_c &=\; \phantom{17}1.67\phantom{0} \;\text{GeV}\qquad\qquad
  m_s &=\; 0.215           \;\text{GeV} \\
  m_t &=\;          172.76\phantom{0}  \;\text{GeV}\qquad\qquad
  m_b &=\; 4.78\phantom{0} \;\text{GeV}
\end{eqnarray*}
\begin{eqnarray*}
  M_W &=\; \phantom{1}80.379\phantom{0} \;\text{GeV} \quad\qquad
  m_e &=\;  0.0005109989461  \;\text{GeV}  \\
  M_Z &=\; \phantom{1}91.1876 \;\text{GeV}  \quad\qquad
  m_{\mu} &=\; 0.1056583745\phantom{000} \;\text{GeV} \\
  M_H &=\; 125.1\phantom{000} \;\text{GeV} \quad\qquad
  m_{\tau} &=\; 1.77686\phantom{00000000} \;\text{GeV}\qquad.
\end{eqnarray*}

\section{Total cross sections and inclusive results of benchmark
  processes}
\label{secinclusiveresults}

For the total cross sections, we restrict ourselves to fully inclusive
results: a complete treatment of these processes with identified
photons in the final state taking into account a sophisticated future
experimental setup like fiducial phase space cuts and selection
efficiencies is beyond the scope of this study. We dedicate
section \ref{differentialresults} to a more detailed investigation of
hard-photon reduced observables by considering cut criteria on the
radiated photon energy for the differential results.

\begin{figure}
  \centering
  \begin{subfigure}{.66\textwidth}
      \includegraphics[width=.45\textwidth]{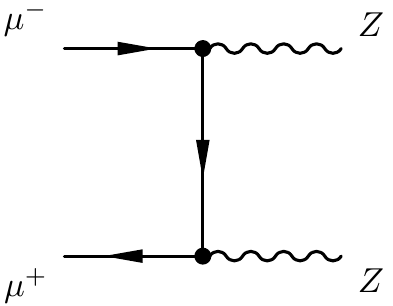}
      \includegraphics[width=.45\textwidth]{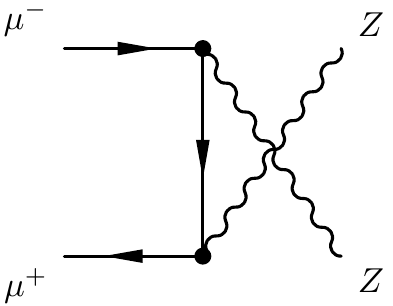}
      \caption{}
      \label{muzz_LO}
  \end{subfigure}
  \begin{subfigure}{.33\textwidth}
    \includegraphics[width=.90\textwidth]{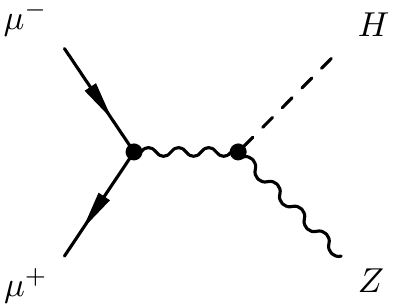}
    \caption{}
    \label{muzh_LO}
    \end{subfigure}
  \\
  \begin{subfigure}{.40\textwidth}
    \includegraphics[width=.90\textwidth]{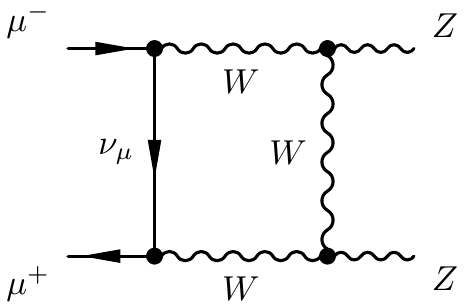}
    \caption{}
    \label{muzz_NLO}
  \end{subfigure}
  \qquad
  \begin{subfigure}{.40\textwidth}
    \includegraphics[width=.90\textwidth]{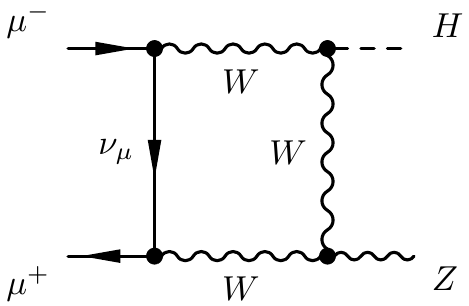}
    \caption{}
    \label{muzh_NLO}
  \end{subfigure}
  \caption{\label{feyndiag_2to2}
    Upper row: tree-level diagrams for the processes
    $\mu^+\mu^-\rightarrow ZZ$ in (a) and to $\mu^+\mu^-\rightarrow HZ$
    (omitting tiny contributions from the muon Yukawa coupling here for
    simplicity) in (b), respectively. Lower row: representative one-loop
    diagrams for the virtual contribution to $\mu^+\mu^-\rightarrow
    ZZ$ in (c) and to $\mu^+\mu^-\rightarrow HZ$ in (d), respectively.}
\end{figure}

For the results presented in
this section we define the relative NLO EW correction
$\delta_{\text{EW}}$ as
\begin{align}
  \delta_{\text{EW}}=
  \frac{\sigma_{\text{NLO}}^{\text{incl}}}{\sigma_{\text{LO}}^{\text{incl}}}-1
\end{align}
where $\sigma^{\text{incl}}_{\text{LO}}$ and
$\sigma^{\text{incl}}_{\text{NLO}}$ are the total inclusive cross
sections at LO and NLO EW, respectively. So, this definition is the
usual NLO K factor subtracted by one.

\subsection{\texorpdfstring{Collider energy scans for cross
    sections of $HZ$ and $ZZ$ production}{Collider energy scans for
    cross sections of {\em HZ} and {\em ZZ} production}}
\label{scans}

The simplest processes where the effects from fundamental EW
higher order perturbative corrections can be understood are neutral
di-boson production processes, as these do not feature final-state,
but only initial-state QED radiation at NLO in $\alpha$. These
processes have a clear kinematical structure at LO dominated by either
$s$-channel or $t/u$-channel/peripheral phase space
configurations. In Fig.~\ref{feyndiag_2to2} we show the tree-level
Feynman diagrams for $\mumut ZZ$ and $ZH$, respectively, in the upper
row. The lower row depicts typical one-loop diagrams, underlining the
fact that at the level of NLO EW corrections these processes are
closely related. The direct contributions associated with muon-Higgs
Yukawa couplings that we always include in the amplitudes in this
work, are of very small sizes and always by far subdominant
(cf. however~\cite{Han:2021lnp} for subtle effects of that coupling at
very high energies).
\begin{figure}
  \centering
  \includegraphics[width=0.9\textwidth]{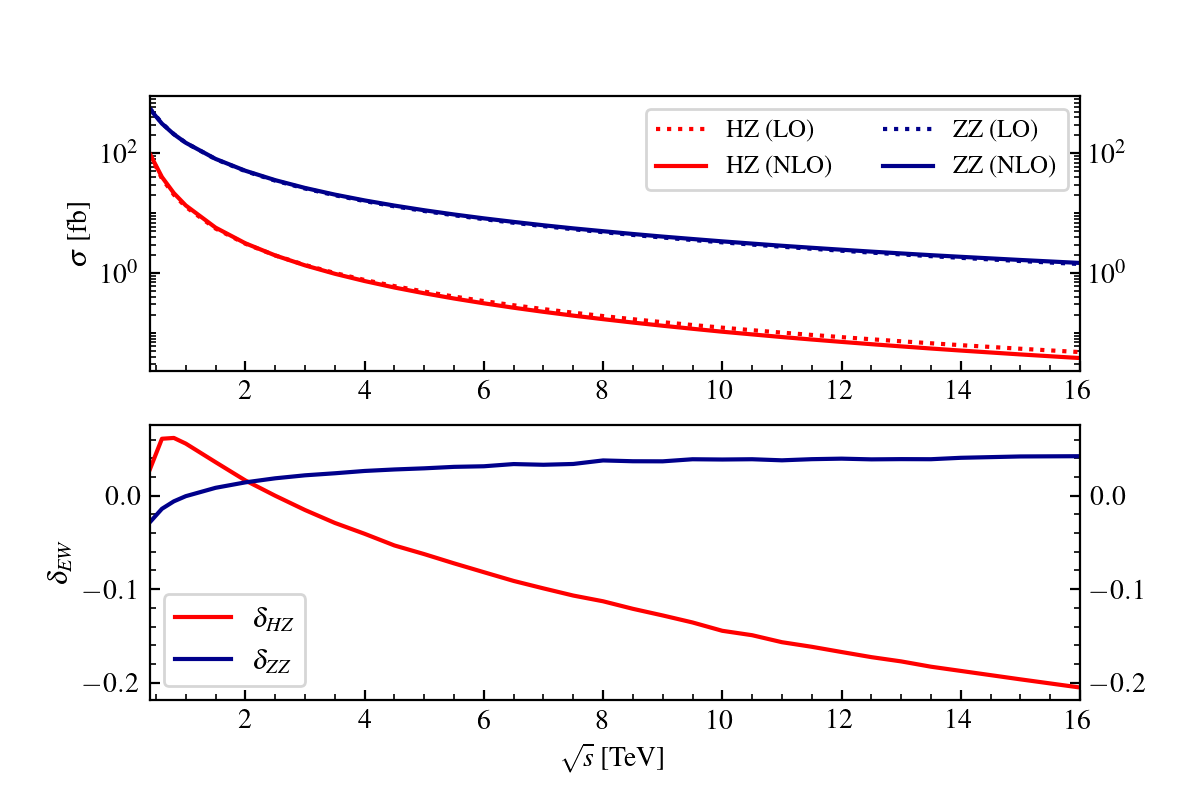}
  \caption{LO and NLO inclusive cross section scans in
    $\sqrt{s}$ for $HZ$ and $ZZ$ production in the upper plot;
    relative NLO correction $\delta_{\text{EW}}$ in the lower
    plot}
  \label{HZaZZ}
\end{figure}
By scans over $\sqrt{s}$ for LO and NLO inclusive cross sections of
$HZ$ and $ZZ$ production at the muon collider, shown in
Fig.~\ref{HZaZZ} (upper plot), we can interprete the global behavior
of the corrections $\delta_{\text{EW}}$ (lower plot) to massive neutral
gauge boson pair production and Higgsstrahlung processes. The
dominant contribution to $ZZ$
production comes from the $t$-channel diagram displayed in
Fig.~(\ref{muzz_LO}), and to $HZ$ production from the $s$-channel
diagram in (\ref{muzh_LO}), respectively. This kinematic
classification of these processes is useful in order to
understand different effects at NLO EW in different kinematic regimes
of invariant masses of external particles.

We first discuss the Higgsstrahlung process. In general, this is very
similar to the corresponding process at an $e^+e^-$ Higgs factory,
where NLO QED corrections are known since a long time, even for
the off-shell process, $e^+e^- \to \mu^+\mu^-H$,
~\cite{Nogueira:1992en,Kniehl:1993ay}, while leading NLO EW corrections
have been calculated in~\cite{Belanger:2002ik}. The on-shell
Higgsstrahlung process $e^+e^- \to ZH$ has recently been also computed
at two-loop~\cite{Song:2021vru}. However, we are
considering the Higgsstrahlung process at the muon collider towards
much higher collider energies, where the threshold is only important
in radiative return events. For this process, $\mu^+\mu^- \rightarrow
HZ$, we observe a large suppression (substantial negative
$\delta_{\text{EW}}$) in Fig.~\ref{HZaZZ} which increases in size from
the peak of the cross section at $\sqrt{s}\sim 0.8$ TeV and which can
be attributed to large virtual effects as we will show in the following.

In order to understand the behavior in the regime of high
center-of-mass energies, in general we can make use of the
approximation of EW Sudakov logarithmic correction factors
for which pioneering works have been done in
\cite{Kuhn:1999nn,Denner:2000jv,Denner:2001gw,Bell:2010gi}. In a
kinematic region,
\begin{align}
  r_{kl}= (p_k+p_l)^2 \sim s \gg M_W^2 
  \label{sudakovlimit}
\end{align}
called the Sudakov limit, with $k$ and $l$ arbitrary external states
carrying (electro-)weak quantum numbers~\footnote{We always assume
the multiplicity of external bosons small enough such that the
condition Eq.~\ref{sudakovlimit} is kinematically valid for all
external legs simultaneously.}, these correction factors
effectively correspond to the purely EW virtual contributions. Note
that in order to restore the full $SU(2)_L\times U(1)_Y$ EW symmetry
and to treat all highly-energetic EW gauge bosons equal, one uses a
fictitious photon mass $\lambda=M_W$. In addition to the EW virtual
corrections, radiative corrections of real photons with transverse
momenta smaller than a cutoff scale at the order of $M_W$ implicitly
are contained in the correction factors.

The high-energy radiative corrections represent form factors in terms of
double and single logarithms of the ratio $r_{kl}/M_W^2$ which are
factorized in the soft and/or collinear limit. If we treat the QED IR
subtraction in an NLO EW computation exactly, thereby keeping the
photon massless as mandated by FKS subtraction, the virtual loop
contributions from massive weak vector boson exchange -- unlike those of
photon exchange -- are still regularized by their masses, i.~e. the EW
scale $M_W$. Therefore, for large $\sqrt{s}$, these contributions are
implicitly contained in the EW next-to-leading logarithmic (NLL)
Sudakov factors. Treating the latter effectively as virtual loop
contributions within the FKS scheme is under the caveat that the IR
singularities from real emission amplitudes and from loop
contributions with photon exchange are regularized at different
(inconsistent in the sense of extracting the Sudakov factors)
scales. This however can not be circumvented in a trivial way if
combining QED FKS subtraction with the Sudakov approximation since for
EW $SU(2)_L\times U(1)_Y$ resummation photon exchange in loop
contributions cannot be treated separately owing to the mixing in the
neutral gauge sector. However, the mismatch of the scales at high
energies has minor numerical effects as it is shown by~\cite{Granata:2017iod}.

For illustration, we show in Fig.~\ref{HZaZZ} the logarithmic
suppression of the NLO cross section for $\mu^+\mu^- \to HZ$ in terms
of the relative correction $\delta_{HZ}$. In the following, we will
show that the effective EW virtual loop contributions of $HZ$ is
quantitatively approximated relatively well enough by the Sudakov
approach.
\begin{figure}
	\centering
	\includegraphics[width=0.9\textwidth]{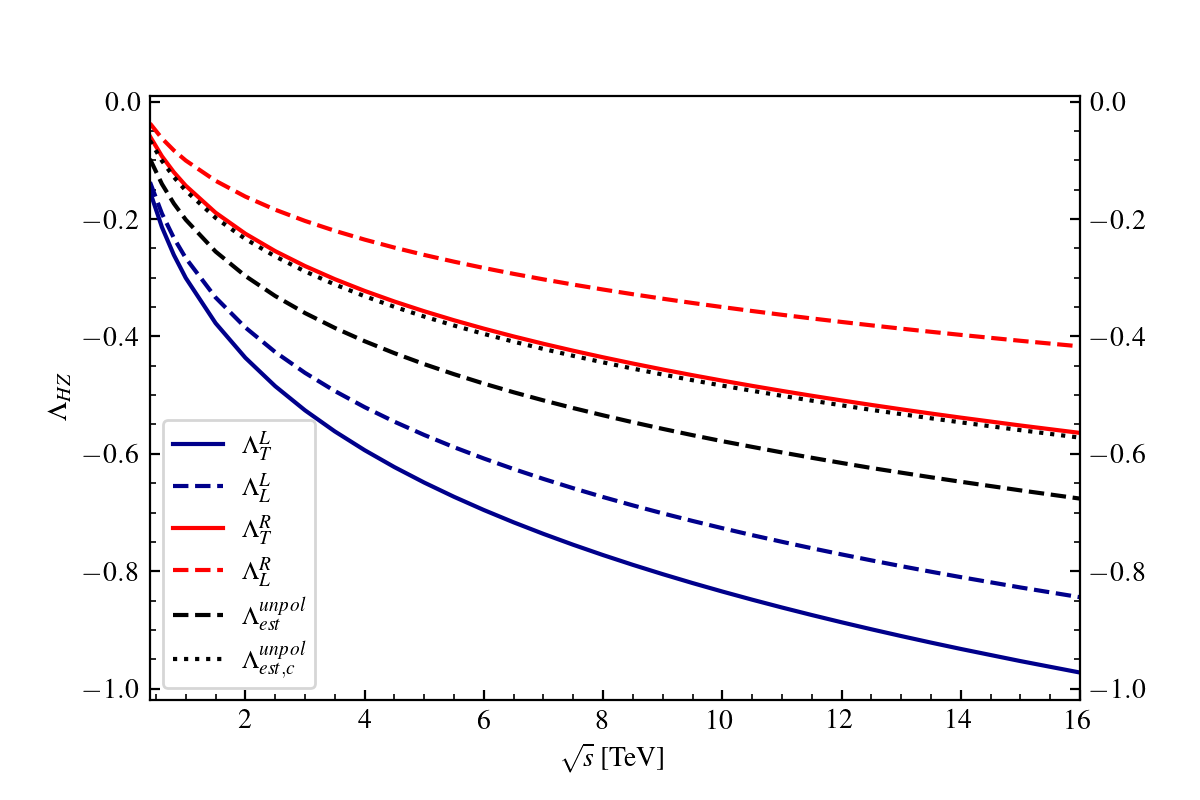}
	\caption{Sudakov factors $\Lambda^{\kappa}_{\lambda}$ for muon chiralities $\kappa=L,R$ and $Z$ polarizations $\lambda=T,L$ and
		estimated unpolarized correction factor
		$\Lambda_{\text{est}}^{\text{unpol}}$ at
		$\theta_{H}=90^{\circ}$ as well as
		$\Lambda^{\text{unpol}}_{\text{est,c}}$ (without angular dependent terms) for $HZ$ production
		at the muon collider as a function of the collider
                energy, $\sqrt{s}$.}
	\label{sudpic}
\end{figure}
To this end, we extract the NLL Sudakov form factor for $HZ$ production
at the muon collider by using analytical results in an analogous way
as for the process $q\bar{q}\rightarrow HZ$~\cite{Granata:2017iod};
the technical details and considerations are explained in
appendix~\ref{HZapprox}, such that we arrive at the estimate of
Eq.~(\ref{sudoverall}). We discuss here only briefly the main
features, for more details cf. appendix~\ref{HZapprox}. This Sudakov
factor as a function of
$\sqrt{s}$ for a fixed solid polar angle $\theta=90^{\circ}$ (note
that the subleading single Sudakov logarithms in
Eq.~(\ref{sudangle}) are functions of the Mandelstam
variables $t$ and $u$) of the Higgs as shown in Fig.~\ref{sudpic}
exhibits large suppressions, in particular for left-handed muons in the
initial state due to the enhanced $SU(2)_L$ weak interaction
coupling.  In a further approach we approximated the unpolarized
correction factor $\Lambda^{\text{unpol}}_{\text{est,c}}$, for which the
angular-dependent part of Eq.~(\ref{sudangle}) is left out,
shown as black dotted curve. This correction factor corresponds to the
amount by which at least the inclusive result is suppressed due to the
virtual loop corrections. According to Eq.~(\ref{sudangle}),
the angular-dependent part of the Sudakov factor is negative, of
subleading logarithmic type and amounts up to $-17\%$ at $\sqrt{s}=16
\text{ TeV}$ for angles $\theta_H$ in the perpendicular plane,
i.~e. close to $90^{\circ}$. For these angles also the Born process is
enhanced according to
\cite{Chanowitz:1985hj,Bohm:2001yx}
\begin{align}
  \left(\frac{d\sigma}{d\Omega}\right)_{\text{Born}}\propto
  \frac{\beta
    M_Z^2}{(s-M_Z^2)^2}\left(\frac{s\beta^2}{8M_Z^2}\sin^2\theta_{H}+1\right)
  \quad.
\end{align}
This angular dependence can be observed as well in the differential
cross sections for the Born case presented in
Sec.~\ref{differentialresults} within Fig.~\ref{thetadist}. The
estimated unpolarized correction factor,
$\Lambda^{\text{unpol}}_{\text{est}}$,  given in
Eq.~(\ref{finaldeltaunpol}) including the angular-dependent
terms, which e.g. for the polar angle $\theta_{H}=90^{\circ}$
decreases the cross section down to $-65\%$, is shown as black
dashed curve in Fig.~\ref{sudpic}.
Both, from the magnitude of $\Lambda^{\text{unpol}}_{\text{est}}$
relative to $\Lambda^{\text{unpol}}_{\text{est},c}$ at this angle as
well as from a similar enhancement behavior at angles around
$\theta_{H}=90^{\circ}$ as for the Born cross section differential in
$\theta_{H}$, the relative inclusive virtual corrections to $HZ$
production can be estimated to be at the order of
$\Lambda^{\text{unpol}}_{\text{est}}$. The counteracting effects of
QED radiation in an inclusive calculation  will be discussed in the
following.

By only including pure NLO QED corrections in the calculation, it can
be shown that the relative correction $\delta_{QED}$ as depicted in
Fig.~\ref{HZQED} is positive and growing with $\sqrt{s}$.
About the contributing NLO parts we can make the
following qualitative statements:
\begin{figure}
  \centering
  \includegraphics[width=0.7\textwidth]{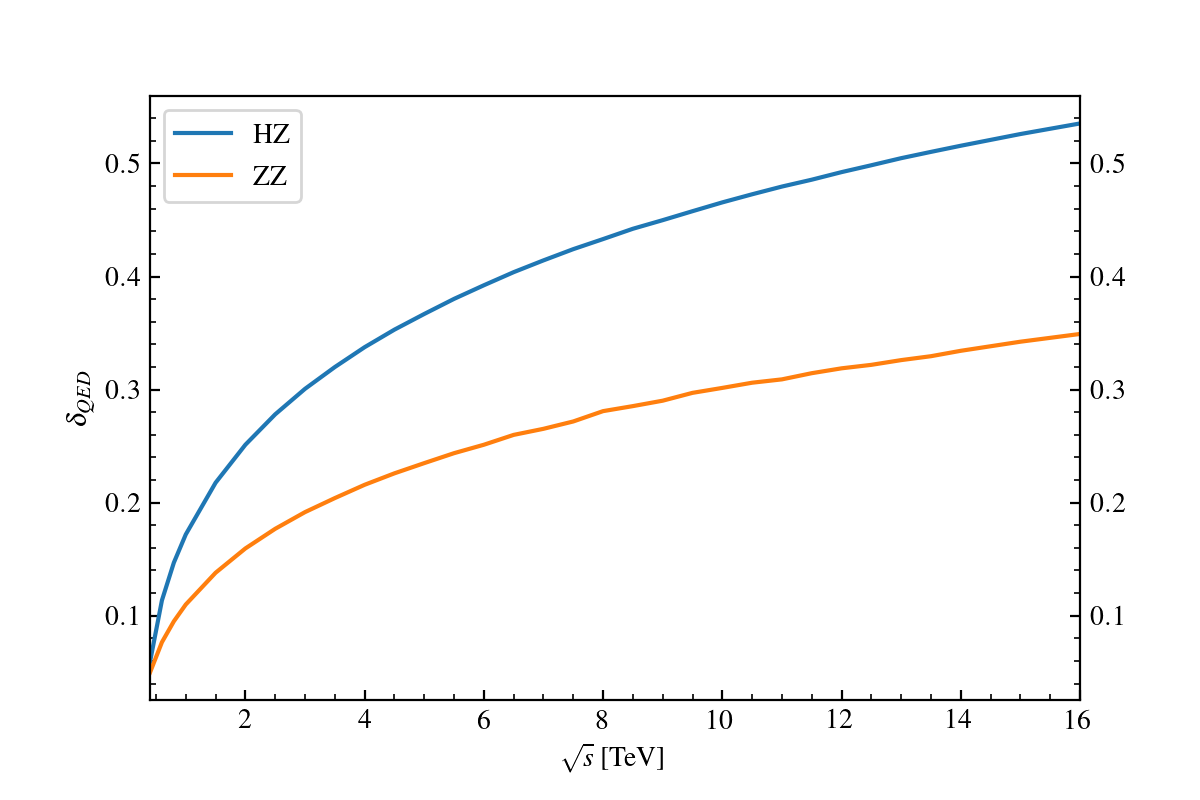}
  \caption{Relative QED corrections
    $\delta_{QED}=\sigma^{\text{incl}}_{\text{NLO,QED}}/\sigma^{\text{incl}}_{\text{LO}}-1$
    to $HZ$ and $ZZ$ production at the muon collider as a function of
    the collider energy, $\sqrt{s}$..}
  \label{HZQED}
\end{figure}
In general, virtual loop amplitudes are supposed to give negative
contributions such that a positive overall NLO correction factor can
be explained by dominating real radiative corrections. In particular,
since the main contributions for $HZ$ production at Born-level come
from the s-channel diagram, large radiative QED corrections at NLO are
expected due to large amplitudes for hard photons radiated in forward
direction, the effect of which is enhanced with growing $\sqrt{s}$.
This explains the bulk of the correction factor to be seen in
Fig.~\ref{HZQED} as the blue curve.
From these two effects, the large negative virtual corrections due to
EW Sudakov logarithms overcompensate the positive QED radiation
effects, resulting in an overall decrease of the cross section of
$\delta_{{HZ}} \sim -20~\%$ at high energies in Fig.~\ref{HZaZZ}. This
magnitude is very reasonable from the general considerations of the
size of Sudakov logarithms and the leading logarithm of
quasi-collinear photon radiation.

For the second process, $\mu^+\mu^- \to ZZ$, we refrain from going into
a detailed discussion about the composition of pure weak and QED parts
of the NLO inclusive corrections since the process is more intricate
from its angular dependence at LO as well as at NLO, due to the
presence of $t$ and $u$ channel because of the Bose symmetry of the
final-state particles. Explicitly, the result for the Sudakov
correction factors of the analogous process $e^+e^-\rightarrow ZZ$
given in \cite{Denner:2000jv} cannot be straightforwardly related
to an estimate for the Sudakov suppression of the integrated result
for the virtual corrections. This is due to the completely different
angular dependence of the Sudakov factor depicted in Fig.~7 within
Ref.~\cite{Denner:2000jv} which is minimal at angles around
$90^{\circ}$ compared to the Born amplitudes with the largest
contributions in forward and backward direction. Moreover, in
agreement with the suppression  $\delta_{QED,ZZ}$ relative to
$\delta_{QED,HZ}$ in Fig.~~\ref{HZQED} we can make the general
statement that the impact of real emission amplitudes with hard photon
radiation on the relative NLO correction is reduced compared to that
of $HZ$ production. This is due to the fact that for $ZZ$ production
large contributions coming from forward scattering of the $Z$ bosons
at high energies are present already at Born-level due to the dominant
t-channel process.
Furthermore, for the real-emission process in general the number of
helicity degrees of freedom of the $Z$ bosons in the final state is
increased relative to the Born process. The latter is non-suppressed
only for transversely polarized $Z$ bosons for opposite
helicities~\cite{Denner:1988tv}. Summing and integrating over all
degrees of freedom in the final state for the real-emission amplitudes
and thereby overcompensating the EW Sudakov effects can explain the
overall positive correction $\delta_{ZZ}$ which can be seen in
Fig.~\ref{HZaZZ}.

\subsection{Total NLO cross sections for multi-boson processes}

In the same way as for our specific processes, $HZ$ and $ZZ$
production, NLO EW corrections can be computed for all other possible
combinations of two, three and four bosons in the final state. In
this section, we present numerical results of the LO and
NLO inclusive cross sections for a large variety of
these processes at $\sqrt{s}= 3$ TeV
in Table~\ref{table3TeV}. Table~\ref{table10TeV} and~\ref{table14TeV}
contain the corresponding results for two- and three-boson production at
$\sqrt{s}= 10$ and $14$ TeV, respectively. We note, that for high
center-of-mass energies as well as high EW boson multiplicities,
fixed-order perturbation theory for the electroweak interactions
become insufficient.
For this reason, we omit the computations for four-boson
production at $\sqrt{s}= 10$ and $14$ TeV which yield meaningful
results only by taking appropriate EW higher order resummation
approaches, e.~g. by soft-collinear effective field theory, into
account. Besides the (fixed-order) NLO EW results, cross sections to
tree-level processes including both collinear resummation of leading
logarithms (LL) of ratios $Q^2/m^2_{\mu}$ and Gribov resummation of
soft radiation to all orders in $\alpha$, as well as  hard-collinear
radiation up to $\mathcal{O}(\alpha^3)$ off the initial state are
included in this section. Their numerical results are achieved by
making use of LL lepton PDFs with its known analytical form presented
in~\cite{Cacciari:1992pz,Skrzypek:1990qs,Skrzypek:1992vk}. For each
corresponding collider energy, these results are displayed in
Table~\ref{table3TeVISR}, \ref{table10TeVISR} and \ref{table14TeVISR},
respectively, where the relative correction $\delta_{\text{ISR}}$ is
defined as
\begin{align}
  \delta_{\text{ISR}} =
  \frac{\sigma_{\text{LO+ISR}}^{\text{incl}}}{\sigma_{\text{LO}}^{\text{incl}}}-1
  \qquad .
\end{align}
We note that in the meantime, also NLL lepton PDFs have become
available~\cite{Frixione:2019lga,Bertone:2019hks,Bertone:2022ktl}
which are already implemented in~\texttt{WHIZARD}; as its
infrastructure within the NLO framework is still in validation,
we do not combine our NLO EW cross
sections with those. In addition, for high-energy muon colliders also
EW PDFs might start to play a phenomenological
role~\cite{Han:2020uid}. Their effects from the resummation of
EW bosons in the initial state are not taken into account here.

We start the discussion with a few general remarks on the LO and NLO EW results.
First of all, we observe that for all processes the absolute value of
the cross section decreases with $\sqrt{s}$ and with the number of
bosons in the final state. Except for pure Higgs final state, these
range from $\sim10^{-4}$ to $\sim10^2$ fb. Because of the size of the
cross sections for the tree-level processes for $\mumut HH$ and
$\mumut HHH$ production which is $\lesssim 2\cdot 10^{-7}$ fb due to
the tiny muon-Higgs couplings at the energies far above the $125$ GeV
threshold, we leave out a detailed discussion on the theoretical
background of the shown $\mathcal{O}(\alpha)$ corrections. The
abnormally  large corrections to cross sections labeled with `$*$' in
in Table~\ref{table3TeV}, \ref{table10TeV} and \ref{table14TeV} at
this fixed order can be related to the fact, that the loop-induced
process is of comparable or even larger size than the formally leading
tree-level process. Hence, the formal NLO correction to the tree-level
process should be rather considered as an interference between the
tree-level and the loop-induced process. The square of the one-loop
amplitude as part of the loop-induced process is beyond the
$\mathcal{O}(\alpha)$ correction predictions considered in this paper
and will be deferred to a future publication. However, although not
relevant for the center-of-mass energies used for the simulations of
this study, for completeness we included the LO and formal NLO $\mumut
HH$ and $\mumut HHH$ numerical cross section results in these tables.

\begin{table}
  \centering
  \small
  \begin{tabularx}{0.8\textwidth}{l|r|r|r}
    $\mumut X, \sqrt{s}=3$ TeV  &
    $\sigma_{\text{LO}}^{\text{incl}}$ [fb]  &
    $\sigma_{\text{NLO}}^{\text{incl}}$ [fb] &
    $\delta_{\text{EW}} $ [\%]\\
    &&&
    \\\hline\hline
    $W^+W^-$       &        $  4.6591(2)\cdot 10^2 $      &   $ 4.847(7) \cdot 10^2 $    & $ +4.0(2) $ \\
    $ZZ$      &       $ 2.5988(1) \cdot 10^1$      &  $ 2.656(2) \cdot 10^1 $    & $ +2.19(6) $ \\
    $HZ$   &     $ 1.3719(1) \cdot 10^0 $   &   $ 1.3512(5) \cdot 10^0 $ & $  -1.51(4)$ \\
    $HH$  &    $ 1.60216(7) \cdot 10^{-7} $
    &  $ 5.66(1)\cdot 10^{-7}~^* $     & $  $
    \\\hline

    $ W^+W^-Z$  &   $ 3.330(2) \cdot 10^1 $       &     $ 2.568(8) \cdot 10^1 $   &   $ -22.9(2) $ \\
    $W^+W^-H$      &     $ 1.1253(5) \cdot 10^0 $    &      $ 0.895(2)\cdot 10^{0} $   &  $ -20.5(2) $ \\
    $ZZZ$       &   $ 3.598(2) \cdot 10^{-1} $     &  $ 2.68(1) \cdot 10^{-1} $   &  $ -25.5(3)$\\
    $HZZ$      &    $  8.199(4)\cdot 10^{-2} $  &   $ 6.60(3)\cdot 10^{-2} $    & { $ -19.6(3) $} \\
    $ HHZ$   & $ 3.277(1) \cdot 10^{-2} $ &  $ 2.451(5) \cdot 10^{-2} $ & $ -25.2(1)$\\
    $ HHH$     &   $ 2.9699(6) \cdot
    10^{-8} $  &   $ 0.86(7) \cdot 10^{-8}~^* $  &  $  $
    \\\hline

    $W^+W^-W^+W^-$   &   $ 1.484(1) \cdot 10^0 $  &    $  0.993(6)\cdot 10^0 $  &   $ -33.1(4) $ \\
    $W^+W^-ZZ$     &   $  1.209(1)\cdot 10^0 $  & $ 0.699(7) \cdot 10^0 $ &  $ -42.2(6) $ \\
    $W^+W^-HZ$      &       $ 8.754(8)  \cdot 10^{-2} $ & $ 6.05(4) \cdot 10^{-2} $ & $ -30.9(5) $\\
    $W^+W^-HH$    &   $ 1.058(1) \cdot 10^{-2} $      &   $ 0.655(5)\cdot 10^{-2} $   &  $ -38.1(4) $\\
    $ZZZZ$     &   $ 3.114(2) \cdot 10^{-3} $    &     $ 1.799(7)\cdot 10^{-3} $   &  $ -42.2(2) $\\
    $HZZZ$     &   $ 2.693(2)\cdot 10^{-3} $    &     $ 1.766(6)\cdot 10^{-3} $   &  $ -34.4(2) $\\
    {$HHZZ$}     &   $ 9.828(7) \cdot 10^{-4} $    &     $ 6.24(2)  \cdot 10^{-4} $   & {$ -36.5(2)$}\\
    $HHHZ$     &   $ 1.568(1) \cdot 10^{-4} $   &      $ 1.165(4) \cdot 10^{-4} $   &  $ -25.7(2) $\\
  \end{tabularx}
  \caption[muon collider NLO EW]{Total inclusive cross sections
    at LO and NLO EW with corresponding relative corrections
    $\delta_{\text{EW}}$, for two-, three- and four-boson
    production at $\sqrt{s}= 3$ TeV. For (*), with dominant
    loop-induced contributions, we refer to the discussion
    in the text.}
  \label{table3TeV}
\end{table}

\begin{table}
  \centering
  \small
  \begin{tabularx}{0.81\textwidth}{l|r|r|r}
    $\mumut X, \sqrt{s}=10$ TeV  &
    $\sigma_{\text{LO}}^{\text{incl}}$ [fb]  &
    $\sigma_{\text{NLO}}^{\text{incl}}$ [fb] &
    $\delta_{\text{EW}}$ [\%]\\
    &&&\\
    \hline\hline
    $W^+W^-$        &       $ 5.8820(2) \cdot 10^1 $  &    $ 6.11(1) \cdot 10^1 $    & $ +3.9(2) $ \\
    $ZZ$        &       $ 3.2730(4) \cdot 10^0$     &    $ 3.401(4)\cdot 10^0 $    & $+3.9(1) $ \\
    $HZ$   &     $ 1.22929(8) \cdot 10^{-1} $  &   $  1.0557(8)\cdot 10^{-1} $ & $ -14.12(7) $ \\
    $HH$ &          $ 1.31569(5) \cdot
    10^{-9} $  &     $ 42.9(4)\cdot 10^{-9}~^{*} $     & $$
    \\\hline

    $ W^+W^-Z$   &   $ 9.609(5) \cdot 10^{0} $      &     $  5.86(4)\cdot 10^0 $   &   $ -39.0(2) $ \\
    $W^+W^-H$           &     $ 2.1263(9)\cdot 10^{-1} $    &      $ 1.31(1)\cdot 10^{-1} $   &  $ -38.4(5)  $ \\
    $ZZZ$        &   $ 8.565(4) \cdot 10^{-2} $  &  $ 5.27(8) \cdot 10^{-2} $   &  $ -38.5(9)$\\
    {$HZZ$}       &    $  1.4631(6)\cdot 10^{-2} $  &  $0.952(6)\cdot 10^{-2} $    & { $ -34.9(4) $} \\
    $ HHZ$   & $ 6.083(2) \cdot 10^{-3} $  &$  2.95(3)\cdot 10^{-3} $ & $ -51.6(5) $\\
    $ HHH$     &   $ 2.3202(4) \cdot 10^{-9} $  &   $ -1.0(2) \cdot 10^{-9}~^{*}  $  &  $ $\\
  \end{tabularx}
  \caption[muon collider NLO EW]{Total inclusive cross sections
    at LO and NLO with corresponding relative correction
    $\delta_{\text{EW}}$ for di- and tri-boson production at
    $\sqrt{s}= 10$ TeV. For (*), with dominant loop-induced
    contributions,  we refer to remarks in the text.}
  \label{table10TeV}
\end{table}

\begin{table}
  \centering
  \small
  \begin{tabularx}{0.83\linewidth}{l|r|r|r}
    $\mumut X, \sqrt{s}=14$ TeV  &
    $\sigma_{\text{LO}}^{\text{incl}}$ [fb]  &
    $\sigma_{\text{NLO}}^{\text{incl}}$ [fb] &
    $\delta_{\text{EW}}$ [\%]\\
    &&&\\
    \hline\hline
    $W^+W^-$        &       $  3.2423(1)\cdot 10^1 $   &     $  3.358(8)\cdot 10^1 $    & $ +3.6(2) $ \\
    $ZZ$       &       $ 1.80357(9) \cdot 10^0$   &     $ 1.872(4)\cdot 10^0 $    & $ +3.8(2)$ \\
    $HZ$   &     $  6.2702(4)\cdot 10^{-2} $  &   $  5.097(6)\cdot 10^{-2} $ & $ -18.7(1) $ \\
    $HH$  &          $ 3.4815(1) \cdot 10^{-10} $  &     $
    217.(2)\cdot 10^{-10}~^{*} $     & $ $
    \\\hline

    $ W^+W^-Z$  &    $ 6.369(3) \cdot 10^{0} $    &        $  3.51(3)\cdot 10^0 $   &   $ -45.0(4) $ \\
    $W^+W^-H$           &     $ 1.2846(6)\cdot 10^{-1} $   &      $ 0.73(1)\cdot 10^{-1} $   &  $ -43.3(9) $ \\
    $ZZZ$        &   $  5.475(3)\cdot 10^{-2} $   &   $  3.06(3)\cdot 10^{-2} $   &  $ -44.2(6)$\\
    {$HZZ$}      &    $  8.754(4)\cdot 10^{-3} $ &    $5.28(3)\cdot 10^{-3} $    & { $ -39.7(4) $} \\
    $ HHZ$    & $  3.668(1)\cdot 10^{-3} $  & $  1.49(1)\cdot 10^{-3} $ & $ -59.4(3)$\\
    $ HHH$     &   $  1.1701(2)\cdot 10^{-9} $  &   $  -0.739(8)\cdot 10^{-9}~^{*} $  &  $  $\\
  \end{tabularx}
  \caption[muon collider NLO EW]{Total inclusive cross sections
    at LO and NLO with corresponding relative correction
    $\delta_{\text{EW}}$ for di- and tri-boson production at
    $\sqrt{s}= 14$ TeV. For (*), with dominant loop-induced
    contributions, we refer to remarks in the text.}
  \label{table14TeV}
\end{table}

For the two specific di-boson processes $\mumut ZZ$ and $\mumut HZ$,
we refer to the detailed discussion within the previous chapter and
highlight the different kinematical effects of these
processes. The $W$ pair production, $\mumut W^+W^-$, despite being
different to $ZZ$ at lower energies due to the interference with the
$s$-channel, is similar at higher energies due to dominant
contributions from $t$-channel diagrams at Born level, and also
initial-state radiation patterns similar to to $\mumut ZZ$. The main
differences are the real-emission amplitudes with photon radiation off
the final state. This can induce semi-collinear effects in $W\gamma$
splittings which increase with the energy scale of the
process. However, this is a minor effect compared to large
contributions from semi-collinear photon radiation off the light
muons in the initial state. Another difference to $ZZ$ is that
the $W$ bosons can have {\em two} longitudinal gauge boson
polarizations corresponding to charged Goldstone bosons in the final
state, for which the $s$-channel process is dominant, but which
is suppressed with~$1/s$~\cite{Beenakker:1993tt,Beenakker:1994vn}. These
considerations may explain the similarity of the relative corrections
$\delta_{\text{EW}}$ of the two gauge boson pair production processes
at high energies $10$ and $14$ TeV compared to their difference at $3$
TeV.

The quantitatively different behavior between gauge boson pair
production $\mumut VV$ and Higgsstrahlung $\mumut HZ$ at the considered
collider energies can be seen between the fixed-order electroweak
correction factors $\delta_{\text{EW}}$ and the resummed ISR
correction factors $\delta_{\text{ISR}}$ in Tables~\ref{table3TeVISR},
\ref{table10TeVISR} and \ref{table14TeVISR}. According to these, the
correction from ISR resummation for $\mumut HZ$ is approximately
twice as big as for $\mumut VV$ at 3 TeV and grows with the
energy. This can be understood from pure kinematics of the Born
processes: Higgsstrahlung is $s$-channel and falls off with $1/s$,
hence ISR induces a radiative return back to the threshold and
enhances the cross section at NLO, while for $\mumut VV$ the
$t$-channel dominates and damps the $1/s$ fall-off by a logarithmic
correction towards $\log(s)/s$. Hence, the radiative return is less
prominent for the EW diboson production. Speaking differently, the
large positive $\mathcal{O}(\alpha\ln (Q^2/m_{\mu}^2))$ logarithms due
to hard collinear photon ISR are enhanced with the collider energy and
induce a boost of the photon recoil system along the beam axis. This
causes a forward scattering of the final state massive $HZ$ system and
thus semi-collinear effects in high energy regions of the phase space.

\begin{figure}
  \centering
  \begin{subfigure}{.40\textwidth}
      \includegraphics[width=.90\textwidth]{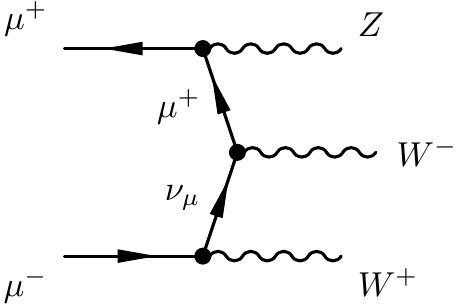}
      \caption{}
      \label{muwwz_LO}
  \end{subfigure}
  \begin{subfigure}{.40\textwidth}
    \includegraphics[width=.90\textwidth]{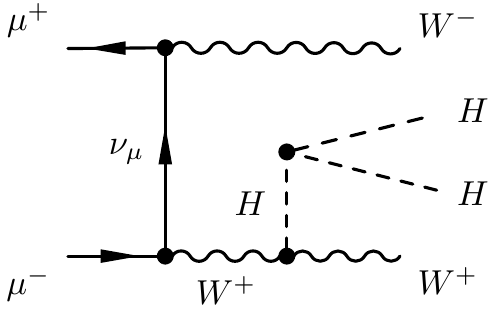}
    \caption{}
    \label{muwwhh_LO}
    \end{subfigure}
  \\
  \begin{subfigure}{.40\textwidth}
    \includegraphics[width=.90\textwidth]{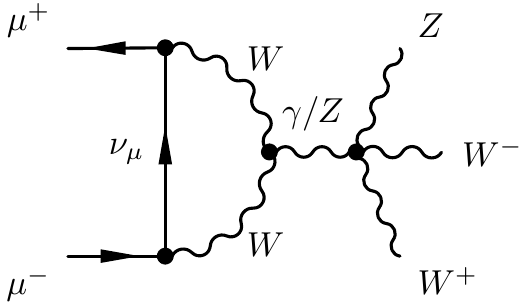}
    \caption{}
    \label{muwwz_NLO}
  \end{subfigure}
  \qquad
  \begin{subfigure}{.40\textwidth}
    \includegraphics[width=.90\textwidth]{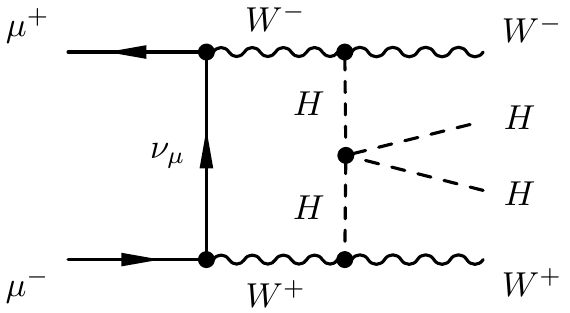}
    \caption{}
    \label{muwwhh_NLO}
  \end{subfigure}
  \caption{\label{feyndiag_2to34}
    Upper row: tree-level diagrams for the processes
    $\mu^+\mu^-\rightarrow W^+W^-Z$ in (a) and to
    $\mu^+\mu^-\rightarrow W^+W^-HH$ 
    in (b), respectively. Lower row: representative one-loop
    diagrams for the virtual contribution to $\mu^+\mu^-\rightarrow
    W^+W^-Z$ in (c) and to $\mu^+\mu^-\rightarrow W^+W^-HH$ in (d),
    respectively.}
\end{figure}

We now turn to processes of higher multiplicities, with three and four
EW gauge and/or Higgs bosons in the final state, again omitting
processes with only Higgs bosons in the final
state. Fig.~\ref{feyndiag_2to34} shows the Feynman diagrams for the
two processes $\mumut W^+W^-Z$ and $\mumut W^+W^-HH$,
respectively. The upper row shows a typical tree-level diagrams, while
in the lower row representative one-loop diagrams are shown, which
include e.g. quartic gauge or Higgs couplings.


For the triple-boson processes at $3$ TeV, we observe a suppression of
$-20\%$ to $-25\%$ of the NLO EW cross section relatively to the LO
result, summarized in Table~\ref{table3TeV}. Compared to the small
absolute value of $\delta_{\text{EW}}$ of the di-boson processes, this
behavior is due to the fact that negative Sudakov logarithm
factors add up for all external states in kinematic regimes where the
Sudakov limit, Eq.~(\ref{sudakovlimit}), is
fulfilled~\cite{Denner:2001gw}. For four-boson final states, this
effect is further enhanced, seen by $\delta_{\text{EW}}$ ranging from
$-26~\%$ to $ - 42~\%$. In addition,
it can be seen from Table \ref{table3TeVISR}, \ref{table10TeVISR} and
\ref{table14TeVISR} that with increasing number of bosons in the final
state the enhancement due to ISR effects decreases. This comes from
the multi-peripheral kinematics of these hard processes, similar to
what was described above for the diboson processes: the radiative
return is not so much pronounced as there is not a single dominating
threshold to return to. Consequently, the dominant contribution to the full
NLO EW correction $\delta_{\text{EW}}$ for three-boson final states
and nearly the complete contribution of $\delta_{\text{EW}}$ for the
four-boson processes is purely due to negative EW virtual final state
correction factors.

\begin{table}
  \centering
  \small
  \begin{tabularx}{0.8\textwidth}{l|r|r|r}
    $\mumut X, \sqrt{s}=3$ TeV  & $\sigma_{\text{LO}}^{\text{incl}}$ [fb] & $\sigma_{\text{LO+ISR}}^{\text{incl}}$ [fb]  & $\delta_{\text{ISR}} $ [\%]\\
    &&&\\
    \hline\hline
    $W^+W^-$       &       $  4.6591(2)\cdot 10^2 $      & $5.303(2)\cdot 10^2$    & $ +13.82(4) $ \\
    $ZZ$        &       $ 2.5988(1) \cdot 10^1$      &  $3.007(1) \cdot 10^1 $    & $ +15.71(4) $ \\
    $HZ$  &     $ 1.3719(1) \cdot 10^0 $   & $1.7868(4)
    \cdot 10^0 $ &  $ +30.24(3) $ \\
    \hline
    %
    $ W^+W^-Z$  &   $ 3.330(2) \cdot 10^1 $     &  $ 3.427(2)\cdot 10^1 $     &   $ +2.90(9) $ \\
    $W^+W^-H$     &     $ 1.1253(5) \cdot 10^0 $    & $1.2052(7)\cdot 10^0$ &        $ +7.10(8) $ \\
    $ZZZ$      &   $ 3.598(2) \cdot 10^{-1} $    &  $3.786(2)\cdot 10^{-1} $  &   $+5.24(8) $\\
    $HZZ$      &    $  8.199(4)\cdot 10^{-2} $  & $ 8.887(5)\cdot 10^{-2} $ &   { $ +8.39(8) $} \\
    $ HHZ$   & $ 3.277(1) \cdot 10^{-2} $ & $3.525(2)\cdot 10^{-2}$ &  $ +7.58(7)$\\
    \hline
    $W^+W^-W^+W^-$   &   $ 1.484(1) \cdot 10^0 $  & $1.465(1)\cdot 10^0$ &     $ -1.3(1) $ \\
    $W^+W^-ZZ$     &   $  1.209(1)\cdot 10^0 $ & $1.187(1)\cdot 10^0$ &  $ -1.8(1) $ \\
    $W^+W^-HZ$     &       $ 8.754(8)  \cdot 10^{-2} $ & $8.742(8)\cdot 10^{-2}$ &  $ -0.1(1) $\\
    $W^+W^-HH$    &   $ 1.058(1) \cdot 10^{-2} $      & $1.076(1)\cdot 10^{-2}$  &   $ +1.7(1) $\\
    $ZZZZ$     &   $ 3.114(2) \cdot 10^{-3} $    & $ 3.139(3) \cdot 10^{-3} $ &     $ +0.8(1) $\\
    $HZZZ$      &   $ 2.693(2)\cdot 10^{-3} $    & $2.730(2)\cdot 10^{-3}$ &    $ +1.4(1) $\\
    {$HHZZ$}     &   $ 9.828(7) \cdot 10^{-4} $    & $10.042(8)\cdot 10^{-4}$   & {$ +2.2(1) $}\\
    $HHHZ$      &   $ 1.568(1) \cdot 10^{-4} $   &  $1.657(1)\cdot 10^{-4}$    &  $ +5.7(1) $\\
  \end{tabularx}
  \caption[muon collider NLO EW]{Total inclusive cross sections
    at LO with LL ISR photon resummation, and including relative correction
    $\delta_{\text{ISR}}$ for two-, three- and four-boson production at
    $\sqrt{s}= 3$ TeV.}
  \label{table3TeVISR}
\end{table}

\begin{table}
  \centering
  \small
  \begin{tabularx}{0.83\textwidth}{l|r|r|r}
    $\mumut X, \sqrt{s}=10$ TeV  & $\sigma_{\text{LO}}^{\text{incl}}$ [fb] &  $\sigma_{\text{LO+ISR}}^{\text{incl}}$ [fb]  & $\delta_{\text{ISR}}$ [\%]\\
    &&&\\
    \hline\hline
    $W^+W^-$       &       $ 5.8820(2) \cdot 10^1 $  & $7.295(7)\cdot 10^1$      & $ +24.0(1) $ \\
    $ZZ$      &       $ 3.2730(4) \cdot 10^0$   & $4.119(4)\cdot 10^0$      & $+25.8(1) $ \\
    $HZ$  &     $ 1.22929(8) \cdot 10^{-1} $ & $1.8278(5)\cdot 10^{-1}$  & $ +48.69(4) $ \\
    \hline
    $ W^+W^-Z$  &   $ 9.609(5) \cdot 10^{0} $      &  $10.367(8)\cdot 10^{0}$    &   $ +7.9(1) $ \\
    $W^+W^-H$     &     $ 2.1263(9)\cdot 10^{-1} $    & $2.410(2)\cdot 10^{-1}$   &  $ +13.3(1)  $ \\
    $ZZZ$     &   $ 8.565(4) \cdot 10^{-2} $  &  $9.431(7)\cdot 10^{-2}$     &  $+10.1(1) $\\
    {$HZZ$}      &    $  1.4631(6)\cdot 10^{-2} $  & $1.677(1)\cdot 10^{-2}$      & { $ +14.62(8) $} \\
    $ HHZ$  & $ 6.083(2) \cdot 10^{-3} $ & $6.916(3)\cdot 10^{-3}$  & $ +13.68(6) $
  \end{tabularx}
  \caption[muon collider NLO EW]{Total inclusive cross sections
    at LO with LL ISR photon resummation and relative correction
    $\delta_{\text{ISR}}$ for two- and three-boson production at
    $\sqrt{s}= 10$ TeV.}
  \label{table10TeVISR}
\end{table}
\begin{table}
  \centering
  \small
  \begin{tabularx}{0.82\linewidth}{l|r|r|r}
    $\mumut X, \sqrt{s}=14$ TeV & $\sigma_{\text{LO}}^{\text{incl}}$ [fb] & $\sigma_{\text{LO+ISR}}^{\text{incl}}$ [fb] &  $\delta_{\text{ISR}}$ [\%]\\
    &&&\\
    \hline\hline
    $W^+W^-$       &       $  3.2423(1)\cdot 10^1 $   & $4.162(4)\cdot 10^1$  &   $ +28.4(1) $ \\
    $ZZ$       &       $ 1.80357(9) \cdot 10^0$   &  $2.288(1)\cdot 10^0$    & $ +26.86(6) $ \\
    $HZ$  &     $  6.2702(4)\cdot 10^{-2} $ & $9.692(3)\cdot 10^{-2}$  & $ +54.57(5) $ \\
    \hline
    $ W^+W^-Z$  &   $ 6.369(3) \cdot 10^{0} $    &   $6.961(6)\cdot 10^{0}$   &   $ +9.3(1) $ \\
    $W^+W^-H$     &     $ 1.2846(6)\cdot 10^{-1} $   & $1.477(1)\cdot 10^{-1}$   &  $ +14.98(9) $ \\
    $ZZZ$      &   $  5.475(3)\cdot 10^{-2} $   &  $6.110(5)\cdot 10^{-2}$  &    $ +11.6(1) $\\
    {$HZZ$}      &    $  8.754(4)\cdot 10^{-3} $ & $10.197(7)\cdot 10^{-3}$ &    { $ +16.49(9) $} \\
    $ HHZ$   & $  3.668(1)\cdot 10^{-3} $ & $4.237(2)\cdot 10^{-3}$ &  $ +15.51(7)$\\
  \end{tabularx}
  \caption[muon collider NLO EW]{Total inclusive cross sections
    at LO with LL ISR photon resummation and relative correction
    $\delta_{\text{ISR}}$ for two- and three-boson production at
    $\sqrt{s}= 14$ TeV.}
  \label{table14TeVISR}
\end{table}

After all, a pattern for the relative NLO EW correction
$\delta_{\text{EW}}$ at 3 TeV for the two- and three-boson production
processes emerges, which is directly related to their kinematical
structure and remains valid for the results at $10$ and $14$ TeV.
The same reasoning as for the diboson processes $\mumut VV$
can be attributed to NLO correction factors with approximately the
same size for $\mumut WWZ, WWH, ZZZ$ and $HZZ$, respectively, which
can be seen at all energies in the tables.
The most reasonable explanation is that the bulk of their NLO
contributions comes from $t$-channel diagrams with at least two gauge
bosons in the final state which induce enhanced Born and real
amplitudes for small scattering angles and high final state
momenta (forward scattering). This is different to the Drell-Yan-like
Higgs- and di-Higgsstrahlung processes, i.~e. $HZ$ and $HHZ$
production, for which the former compared to $WW/ZZ$ and the latter
compared to $WWH/WWZ/ZZZ/ZZH$ have distinct $\delta_{\text{EW}}$ of
$-15~\%$ to $-20~\%$ in table \ref{table10TeV} and
\ref{table14TeV}. This observation can be related to the behavior of
$\delta_{{HZ}}$ compared to $\delta_{{ZZ}}$ with $\sqrt{s}$ in
Fig.~\ref{HZaZZ} for which a detailed study is given in section
\ref{scans}. Concluding, the kinematical structure, either $t$- or
$s$-channel, of the dominant Born process has a decisive impact on the
relative size of NLO EW corrections to inclusive cross sections.

\section{Differential distributions for NLO EW corrections}
\label{differentialresults}

\begin{figure}
	\centering
	\includegraphics[width=0.49\textwidth]{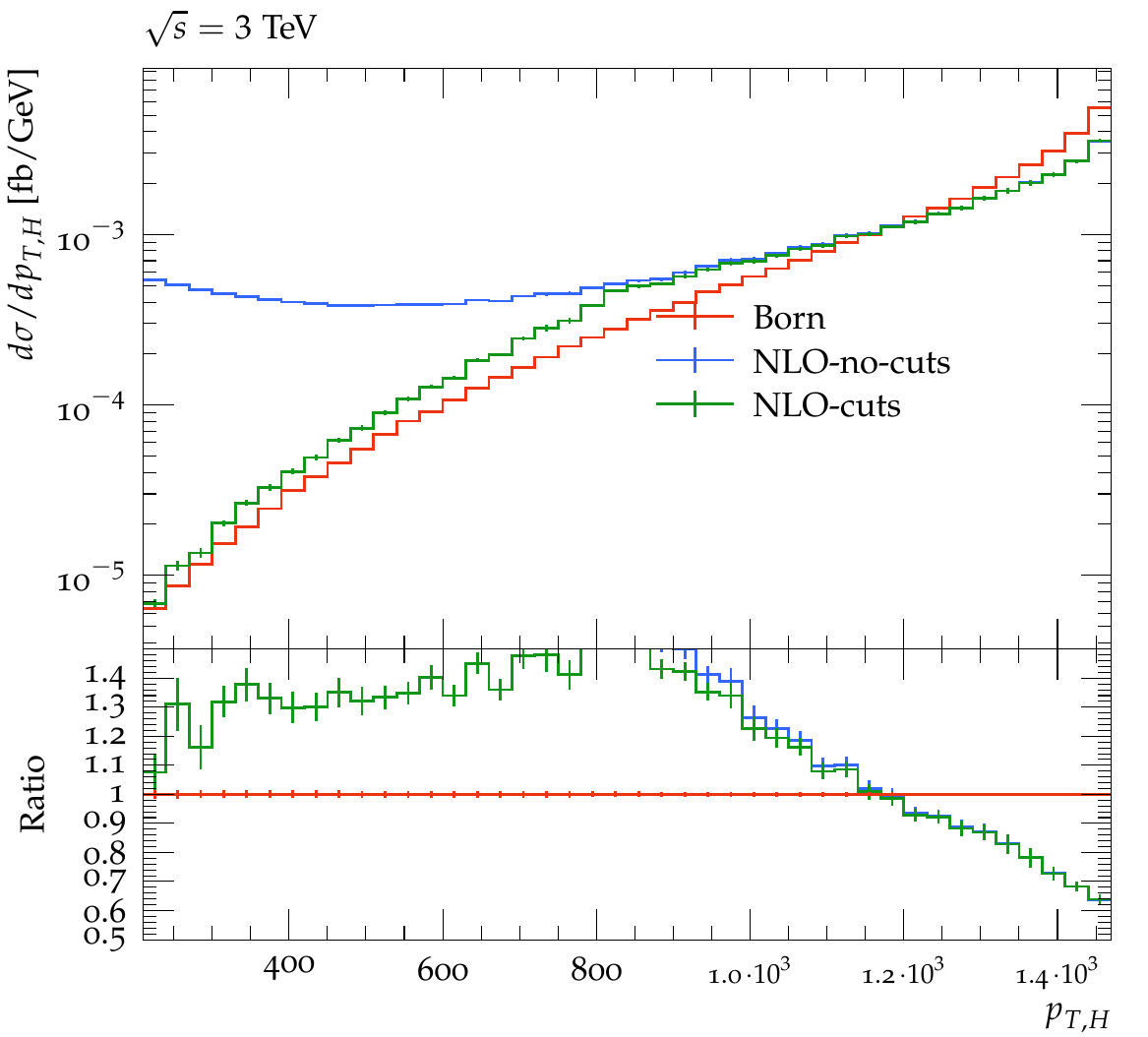}
	\includegraphics[width=0.49\textwidth]{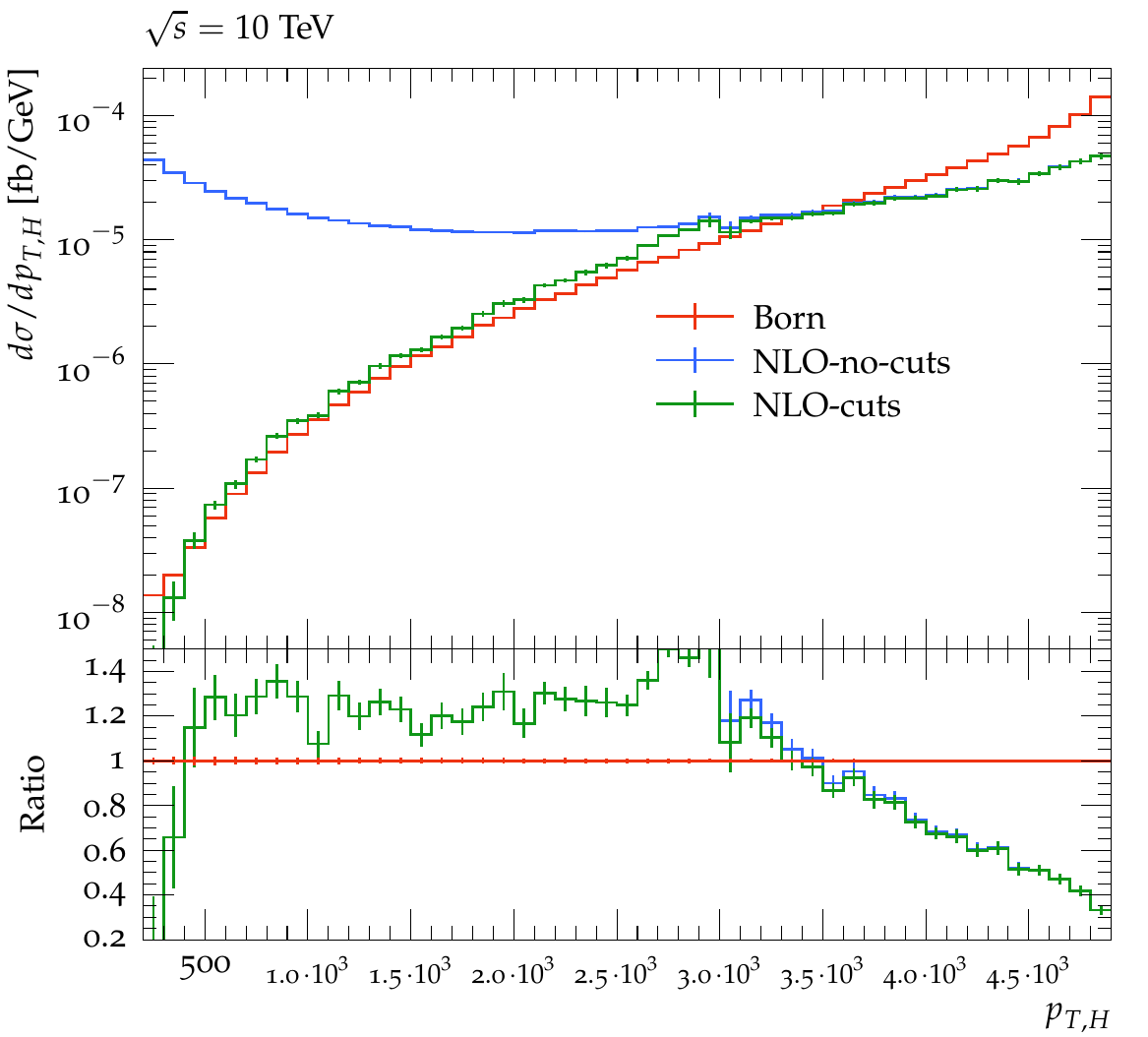}
	\includegraphics[width=0.49\textwidth]{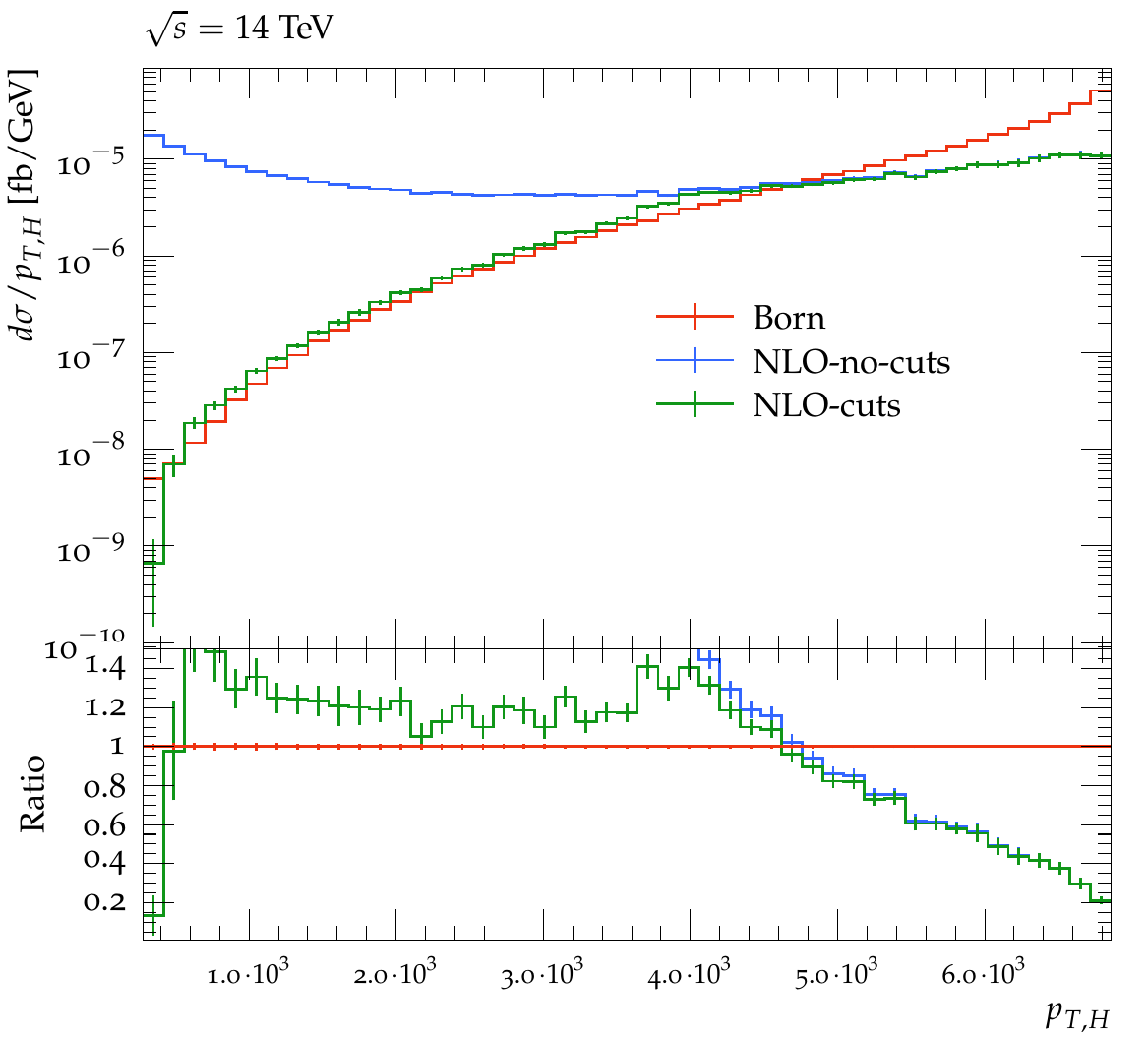}
	\caption{Higgs transverse
		momentum distributions, $d\sigma/dp_{T,H} (\mu^+\mu^-
                \rightarrow HZ)$, at $\sqrt{s} = 3$, $10$ and $14$
                TeV, respectively.}
	\label{ptdist}
\end{figure}

\begin{figure}
	\centering
	\includegraphics[width=0.49\textwidth]{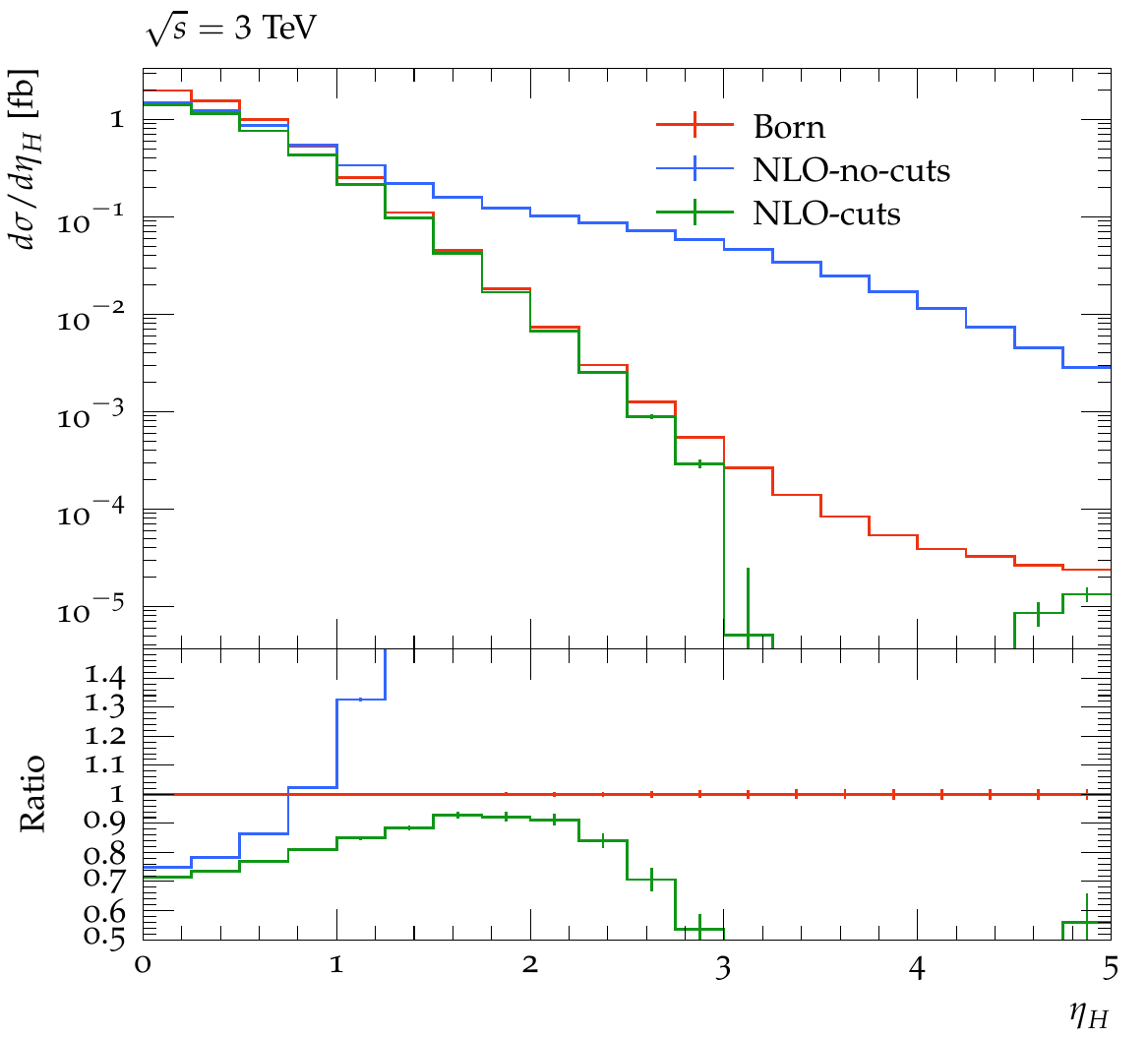}
	\includegraphics[width=0.49\textwidth]{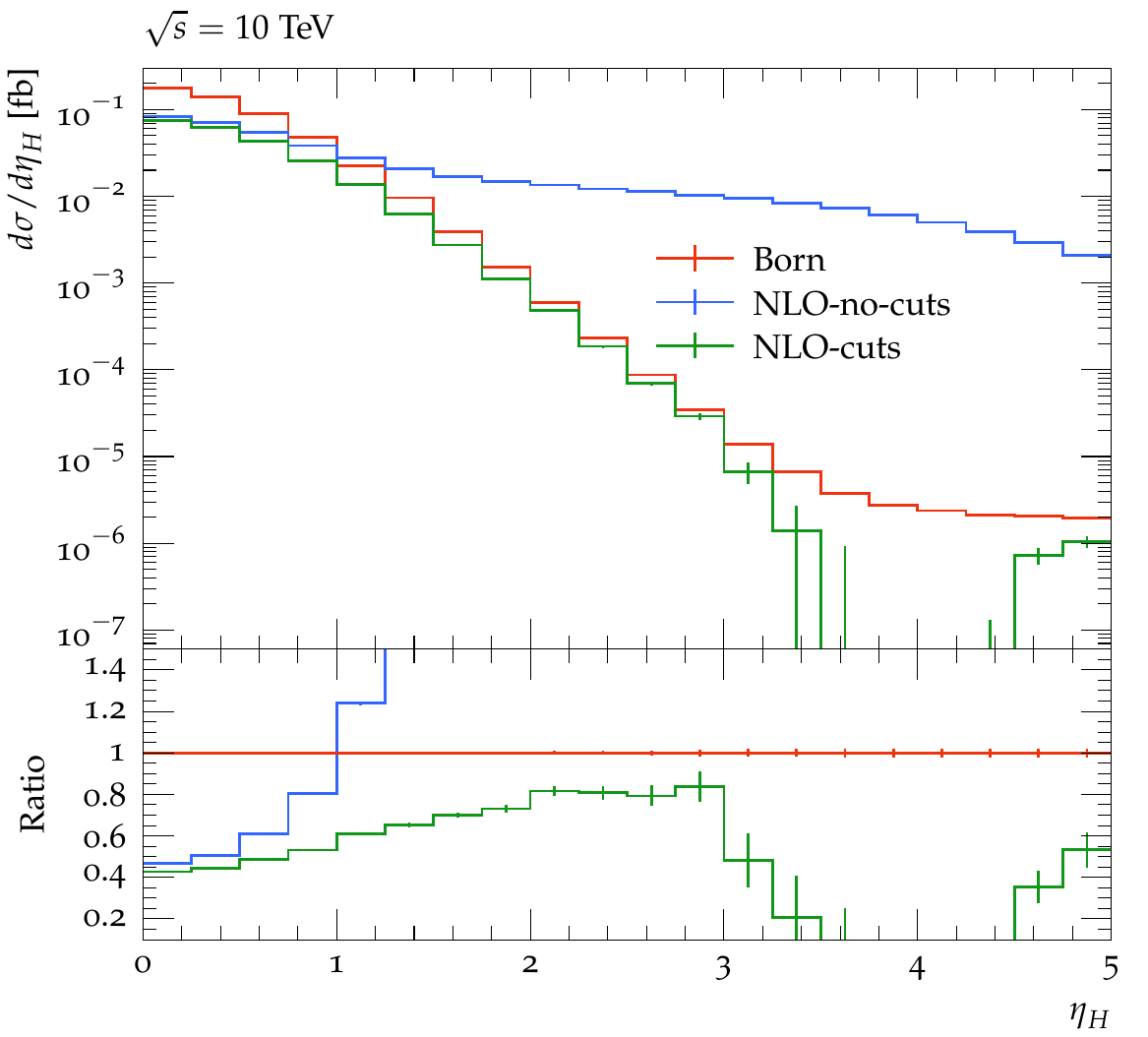}
	\includegraphics[width=0.49\textwidth]{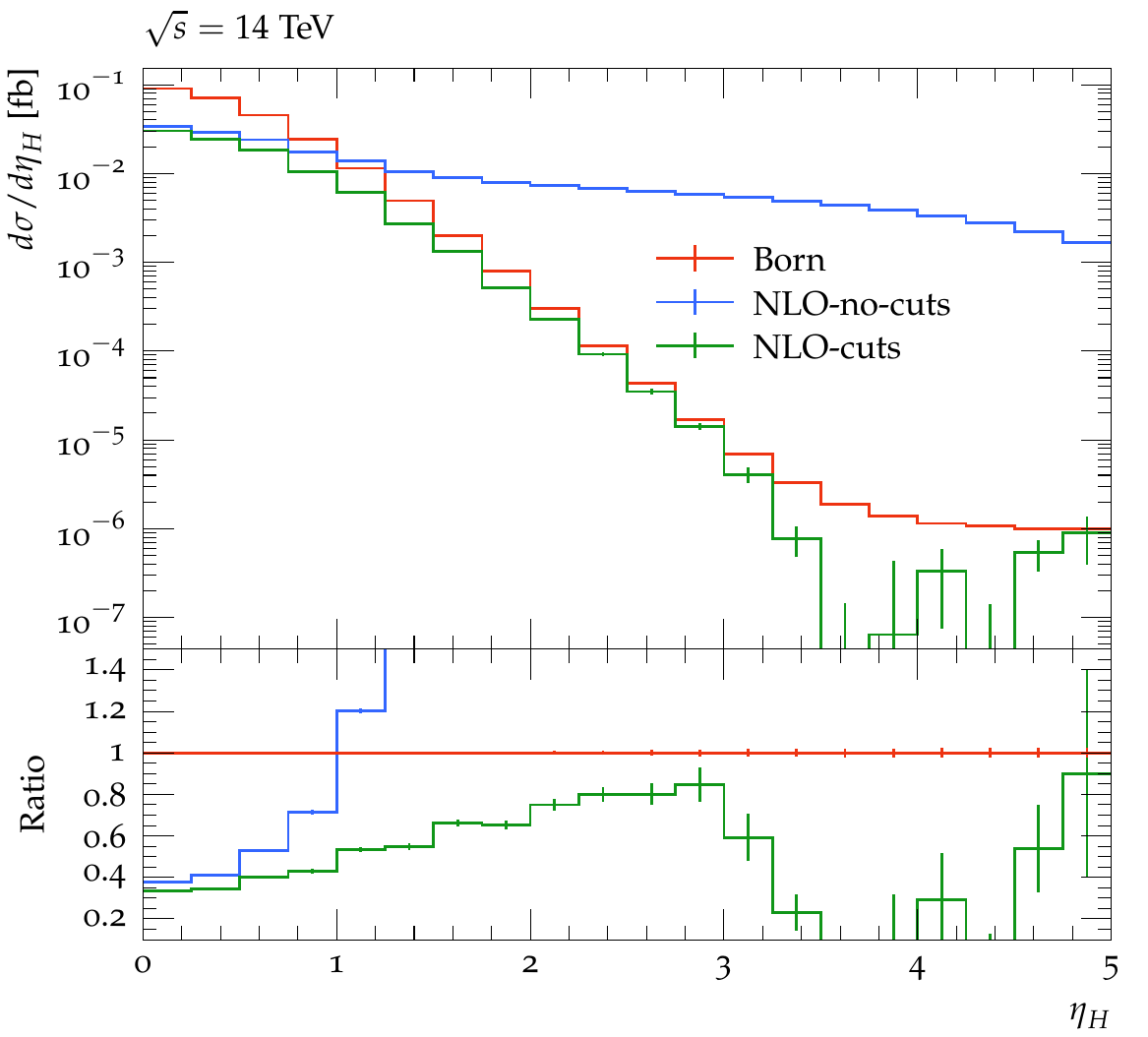}
	\caption{Higgs pseudorapidity distributions,
          $d\sigma/d\eta_{H} (\mu^+\mu^- \rightarrow HZ)$,
          at $\sqrt{s} = 3$, $10$ and $14$ TeV, respectively.
          The distributions are symmetric in $\eta_H$.}
	\label{etadist}
\end{figure}

\begin{figure}
	\centering
	\includegraphics[width=0.49\textwidth]{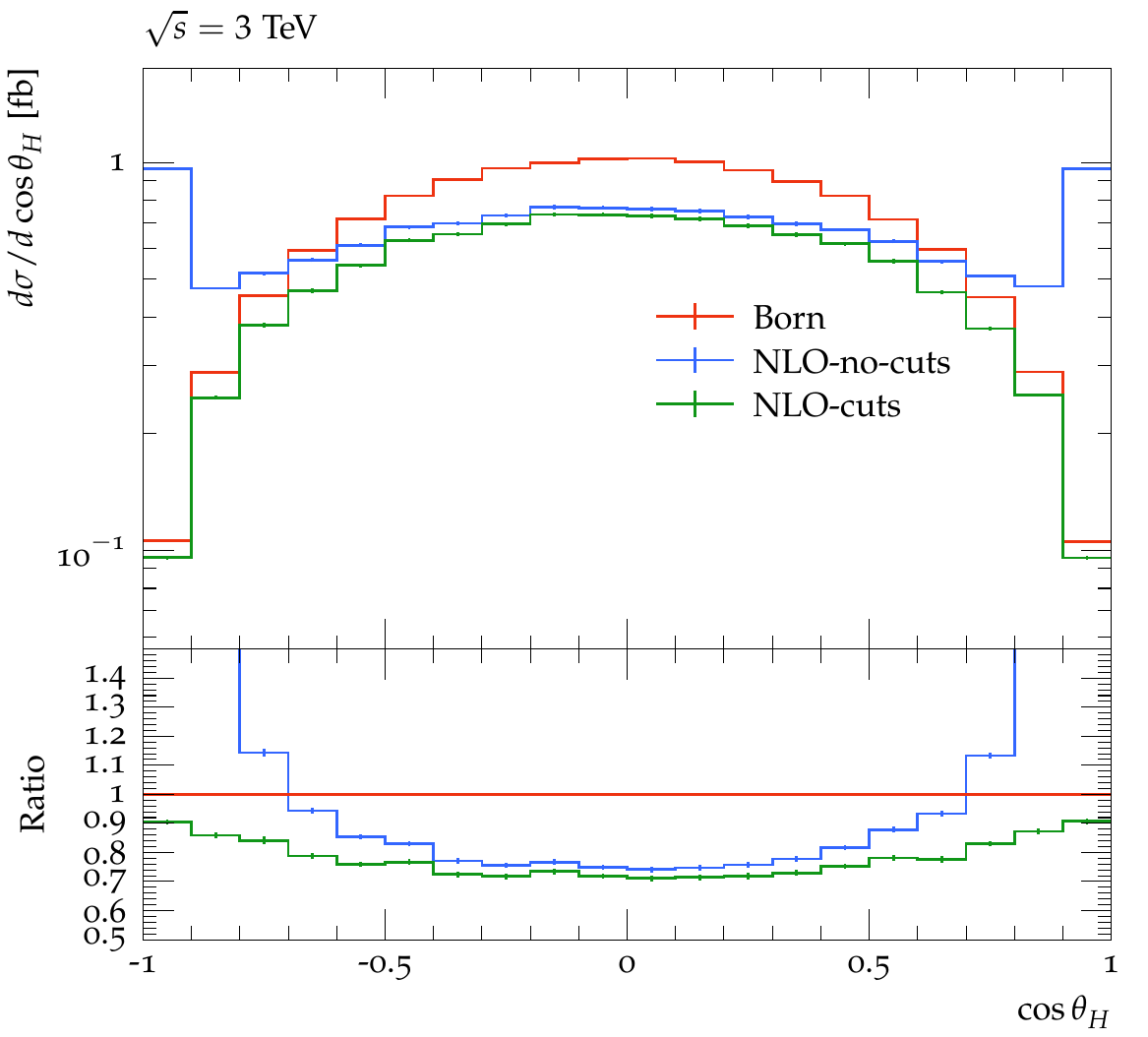}
	\includegraphics[width=0.49\textwidth]{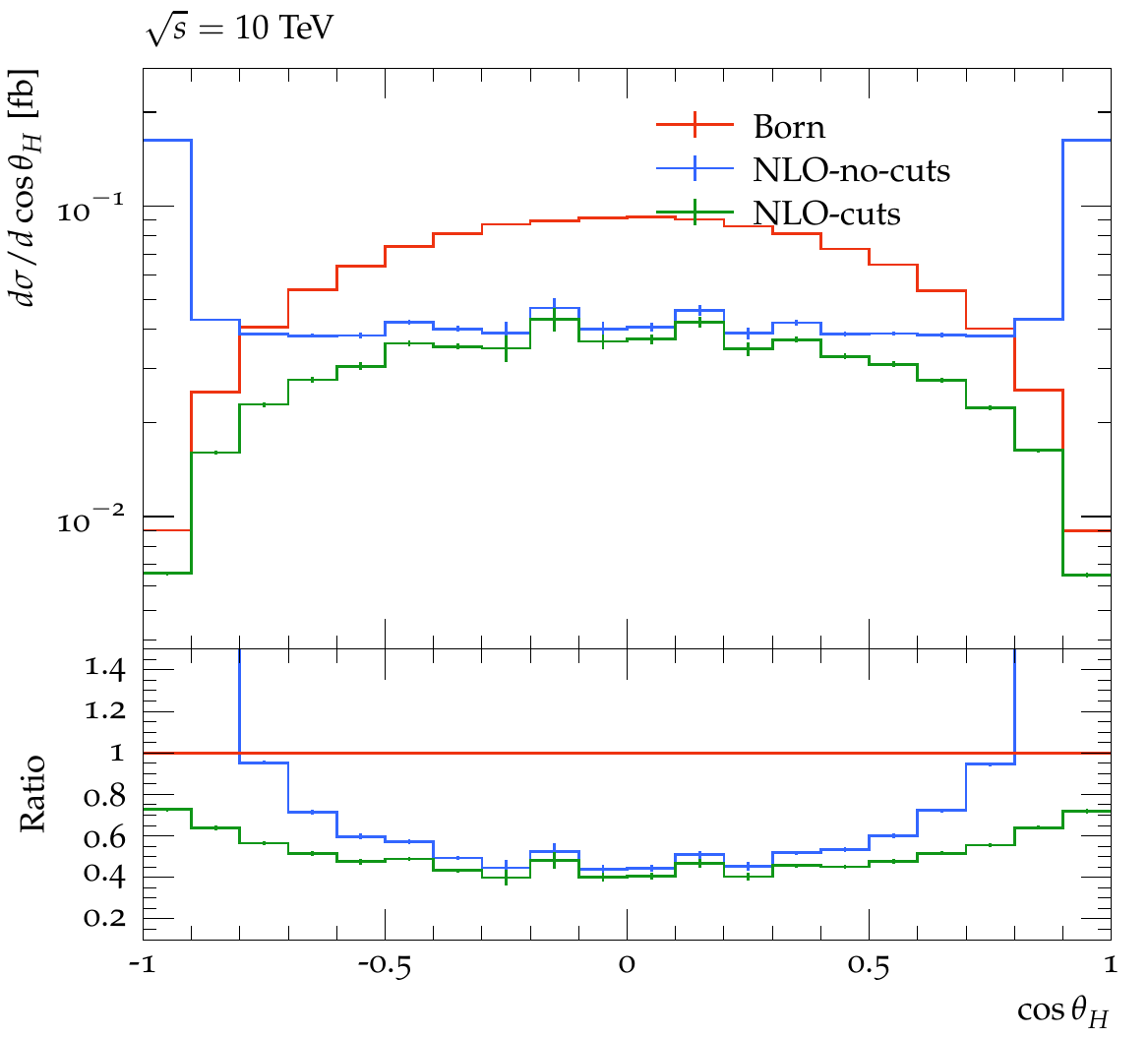}
	\includegraphics[width=0.49\textwidth]{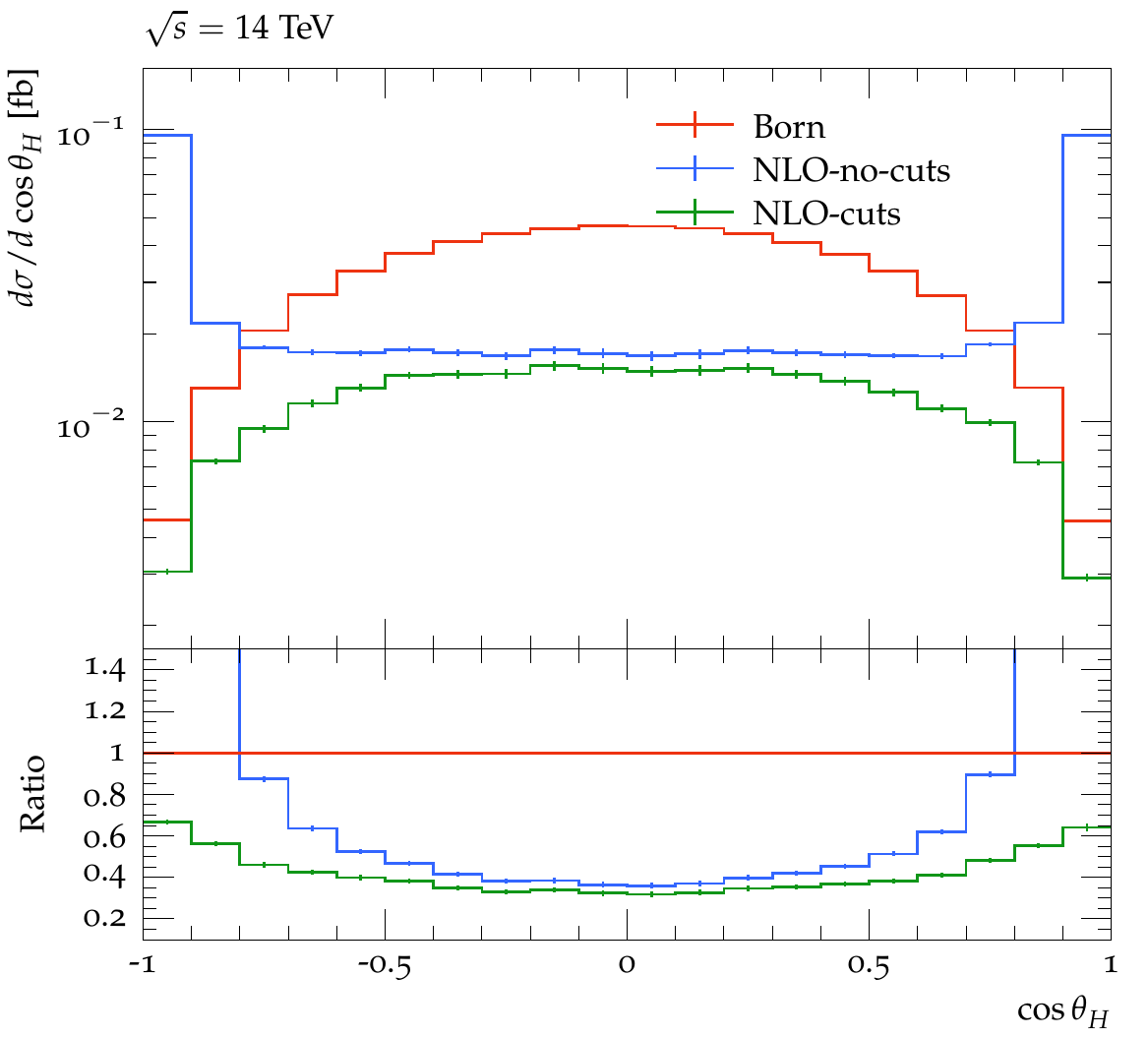}
	\caption{Differential Higgs polar angle distributions,
          $d\sigma/d\cos\theta_{H}(\mu^+\mu^- \rightarrow HZ)$ at
          $\sqrt{s} = 3$, $10$ and $14$ TeV, respectively.}
	\label{thetadist}
\end{figure}

In order to give an overview on the impact of NLO EW
corrections on differential observables, we produce differential
distributions for the process $\mu^+\mu^-\rightarrow HZ$ at $\sqrt{s}
= 3$, $10$ and $14$ TeV, respectively, for different Higgs observables
which are displayed in Fig.~\ref{ptdist}, \ref{etadist} and
\ref{thetadist}, respectively. Obviously, these are fixed-order NLO
differential distributions which for realistic physics simulation
would require a proper matching to QED parton showers in order to
describe all of the electromagnetic activity in the event, which we do
not attempt in this paper. However, we do investigate the effects of EW
corrections on  observables for which cuts on the fiducial phase space
are imposed. In particular, for this case, phase space points with hard photons
exceeding a certain energy are considered as observable photons and
hence are discarded in the analysis. This is along the lines of
typical experimental analyses at high-energy lepton colliders like
ILC~\cite{Berggren:2022}. In order to visualize the impact
of this phase-space cut, we show -- together with the corresponding
Born observable -- two curves for the NLO observables, one for the
case that no cuts are imposed on photon radiation, called
`NLO-no-cuts', and a veto on (very) hard photons,
\begin{align}
  E_{\gamma}<0.7\cdot \sqrt{s}/2,
  \label{photoncut}
\end{align}
which we dub `NLO-cuts' (as there is no QED radiation at the level of
the Born process, such a cut is trivial at LO).

We show distributions for cross sections differential in the Higgs
transverse momentum, $p_{T,H}$, for the three different center-of-mass
energies, 3, 10, and 14 TeV, respectively, in Fig.~\ref{ptdist}. In
these plots we see that the differential K factor, i.e the ratio of
NLO over Born differential cross section is mostly constant for low
$p_{T,H}$ values, reaches a maximum of roughly 1.5-1.6 and then
decreases steeply for $\sqrt{s}/4 \lesssim p_{T,H} \lesssim
\sqrt{s}/2$. In this part of the $p_{T,H}$ range, the curves
`NLO-no-cuts' and `NLO-cuts' almost coincide, and the decrease of the
ratio is steeper the larger the collider energy is, i.~e. the
differential K factor drops to $\sim0.6$ for $\sqrt{s}=3$ TeV,
$\sim0.4$ for $\sqrt{s}=10$ TeV and $\sim0.2$ for $\sqrt{s}=14$ TeV,
respectively.

Again, the origin of this large negative corrections can be traced back
to EW Sudakov logarithmic factors in the form of $\log^2
\left[p_{HZ}^2/M_W^2\right]$, which grow with the invariant mass of
Born $HZ$ Higgsstrahlung system. Obviously, this behavior gets enhanced
the larger the center-of-mass energy of the process is.

The cut on the photon energy influences the differential distributions
in regions which are kinematically not accessible at Born level and
hence receive so-called huge (differential) K factors. This happens in
the region where the Higgs boson has rather small transverse momentum,
$p_{T,H} \lesssim \tfrac12 \; p_{T,H}^{\text{max.}}$, as it recoils at Born
level against the $Z$. This region is then filled by hard photon
radiation at NLO; the veto of Eq.~(\ref{photoncut}) on such hard
radiation reduces the differential K factors to moderate values of
$\sim1.2$. These radiative tails enhance especially the lowest bins in
the $p_{T,H}$ distribution by two, or for $\sqrt{s}=14$ TeV even three
orders of magnitude. Using hadron-collider language, such events would
rather fall into exclusive $ZH$ plus $n$ "jet" bins, $\mumut ZH + n
\gamma$. As for jet vetoes at the LHC, for such hard photon radiation,
a QED resummation is necessary to give a reliable estimate on the
theoretical uncertainty of the prediction in these parts of the phase
space.

Next, we turn to the Higgs pseudorapidity distribution,
$d\sigma/d\eta_{H} (\mumut HZ)$. We make use of the fact that these
distributions are symmetric with respect to the central axis of the
detector, and so we depicted distributions as a function of the
modulus of the Higgs pseudorapidity $|\eta_{H}|$ for the proposed collider
energies in Fig.~\ref{etadist}. First of all, we note that the most
significant deviation between the full and the "vetoed" NLO
distributions in each of the plots is at large pseudorapities, where
the Higgs boson recoils against a hard photon emitted collinear to the
beam axes. In that regime, real matrix elements are drastically
enhanced. On the other hand, for small $|\eta_{H}|$, real photon
radiation for the NLO distributions is suppressed and virtual effects
play a much more significant role. For $|\eta_{H}|=0$, the
Higgs is radiated in the plane perpendicular to the beam axis
such that the value of the differential K factor reduced by one,
$K - 1$, is directly comparable to the Sudakov suppression factor
$\Lambda^{\text{unpol}}_{\text{est}}$ of Eq.~(\ref{finaldeltaunpol});
this factor is shown in
Fig.~~\ref{sudpic}. In fact, the relative deviation of the NLO
differential cross section for which cuts are applied of the Born one
agrees with $\Lambda^{\text{unpol}}_{\text{est}}$ in the first bin for
all the shown plots at the level of a few percent.

Almost the same physics like in the pseudorapiditiy distributions is
encoded in the differential distributions for the Higgs polar angle,
$d\sigma/d\cos \theta_{H} (\mumut ZH)$. However, the polar angle
distributions are much more common for Higgs studies at lepton
colliders. We show the Higgs polar angle distributions for our three
different collider energies of $3$, $10$ and $14$ TeV, respectively,
in Fig.~\ref{thetadist}. Again, one observes that the bulk of the
Born contribution is located at the central part of the detector around
$\theta_{H}=90^{\circ}$. This can be understood from the $\sin^2
\theta_{H}$ dependence of amplitudes with longitudinally polarized $Z$
bosons which are enhanced by $s/M_Z^2$ compared to transversal
ones~\cite{Bohm:2001yx}. By comparing the curves labelled
"NLO-no-cuts" and "NLO-cuts" one sees again that the photon veto,
Eq.~(\ref{photoncut}), cuts out the collinearly strongly enhanced
emission along the beam directions. For angles in the central part of
the phase space, the cut has only a minor effect and the two curves
deviate only at a few percent. As in the case for the central
description in terms of the pseudorapidity, $\eta_{H} \sim 0$,
the Sudakov factor $\Lambda^{\text{unpol}}_{\text{est}}$ can be found to
be an accurate approximation for $d\sigma/d\cos\theta_{H}$ at
$\theta_{H}= \pi/2$, especially for high collider energy and when
applying a hard photon veto.

It would be interesting to also study differential distributions for
production processes with two, three and four electroweak gauge
bosons. However, this chapter served mostly as a proof of principle
and we leave such dedicated studies for future
publications~\cite{NLO_multEW}.


\section{Conclusions and Outlook}

In this paper, we have presented the -- to our knowledge -- first
systematic set of calculations of NLO EW corrections to the production
of two, three and four electroweak gauge and Higgs bosons at a future
high-energy muon collider at three different collider energies of 3,
10 and 14 TeV. Such a collider has recently regained a lot of interest
not as a Higgs factory, but as a machine at the utmost energy
frontier, particularly along the lines of the US Snowmass Community
Study 2021. In order to map out the full physics potential of such a
collider, especially with respect to the discovery reach for new
physics, it is crucial to have precision predictions for SM processes
at hand. We studied a quite extensive list of processes including two,
three and four Higgs and/or electroweak vector bosons at fixed NLO EW
as well as with QED ISR leading-logarithmic collinear
resummation. Processes with only Higgs bosons in the final state are
special, as they are highly suppressed at tree level due to the tiny
muon Yukawa coupling; the tree-level matrix element is of the same
order or even smaller than the one-loop matrix element, such that the
normal loop-wise expansion is not meaningful here. For the other
multi-boson processes, $\mumut V^n H^m, n+m \leq 4, n \neq 0$, the
results presented in this work reflect two significant features of the
higher order EW corrections: on the one hand, large negative virtual
corrections from EW Sudakov double (and single) logarithms, and on the
other hand large collinear logarithms from initial-state photon
radiation. Generically, for all processes the negative EW Sudakov
factors overcompensate for the enhancement of real contributions from
hard photon ISR with the exception of the diboson processes $\mumut
VV$.

The suppression of NLO EW inclusive cross sections with respect to
the LO results range down from $-20\%$ for three bosons at 3 TeV
and to about $-60\%$ for three bosons at 14 TeV. Furthermore,
Higgs- and multi-Higgsstrahlung processes exhibit much smaller
suppression rates compared to other multi-boson processes at high
energies. This is due to their $s$-channel dominated kinematical
structure which allows the resummed ISR collinear radiation a much
stronger enhancement from radiative return.

We also presented results for NLO fixed-order differential
distributions, as a showcase for the process $\mumut HZ$. For this
process, we did an extensive comparison of such differential
distributions to the pure EW NLL Sudakov factor approximation,
and find very good agreement, e.g. the suppression of $d\sigma/d\cos
\theta_{H}$ in distributions for Higgs polar angles perpendicular to
the beam axis. These comparisons also include angular-dependent
(subleading, i.e. single-logarithmic) Sudakov factors. These factors
even more accurately describe the suppression of the distributions if
very hard photon radiation is vetoed by a cut on the radiated photon
energy of $E_{\gamma}<0.7\sqrt{s}$. This regime of the phase space
should be considered as part of the exclusive photon "jet" bins,
$\mumut HZ + n \gamma$.

There are several further roads to pursue from here:
for a universal treatment of collinear ISR effects in the NLO EW
calculation next-to-leading logarithmic (NLL) collinear initial-state
resummation must be applied by factorizing lepton PDFs at this
accuracy level. This would allow to systematically combine the two
different types of corrections properly. In our simulation framework,
this is work in progress and will be available in the future. Such a
theoretical description mandates treating the initial-state muons as
massless. Care has then to be taken when effects from Higgs radiation
off massive muon lines are taken into account for multi-Higgs or
multi-Higgstrahlung processes. One class of electroweak processes that
we have not considered so far, are vector-boson fusion (VBF) or
vector-boson scattering (VBS) like processes, which also are of high
phenomenological importance. Such processes have to be treated as well
with great care, as the quasi-collinear forward neutrinos (for charged
currentss) and muons (for neutral currents) can only be kinematically,
but not conceptually distinguished from decays of single vector
bosons. This is beyond the scope of this paper. Finally, in order to
be able to study detector effects and systematic uncertainties,
unweighted event samples fully matched to initial- and final state
photon showers (or at very high energies even EW showers) have to be
available. We leave this as well for future work.


\acknowledgments

For stimulating discussions on muon colliders and NLO calculations, we
would like to thank Tao Han, Yang Ma and Keping Xie.
Furthermore, we give thanks to Adrian Signer for useful advices on QED
fixed order computations associated with massive initial state
leptons, and to Stefan Kallweit for many helpful details for using
\texttt{RECOLA} for electroweak corrections. We thank Jonas Lindert
for providing corresponding \texttt{OpenLoops} process libraries. We
also appreciate valuable discussions with Stefano Frixione on
theoretical description of initial-state radiation, as well as Mikael
Berggren on the treatment of photons in experimental lepton collider
analyses.  PB, JRR, and PS acknowledge the support by the Deutsche
Forschungsgemeinschaft (DFG, German Research Association) under
Germany's Excellence Strategy-EXC 2121 "Quantum Universe"-3908333. WK
was supported in part by the Deutsche Forschungsgemeinschaft (DFG,
German Research Foundation) under grant 396021762 - TRR 257. This work
is also been funded by the Deutsche Forschungsgemeinschaft (DFG,
German Research Foundation) - 491245950.

\newpage
\appendix

\section{Validation for massive lepton-initiated process setup}
\label{validation}
For the validation of the \texttt{WHIZARD+RECOLA} setup for simulating
cross sections of lepton collider processes at NLO~EW we used the
input parameters and reference results of \cite{Sadykov:2020dgm} for
the process $e^+e^-\rightarrow HZ$ with unpolarized beams and massive
initial state. A comparison of these checks for different collider
energies, i.~e. $250$, $500$ and $1000$ GeV, is shown in
table~\ref{eeHZtable} below. Note that the shown precisions are not
theory uncertainties but just Monte Carlo integration errors which
have been chosen to find well enough agreement between the two
programs at both LO and NLO.

\begin{table}[h]
  \centering
  \small
  \begin{tabularx}{0.9\linewidth}{l|r|r|r|r|r|c}
    &\multicolumn{2}{c|}{\texttt{MCSANCee}\cite{Sadykov:2020dgm}}&\multicolumn{3}{c|}{\texttt{WHIZARD+RECOLA}}&\multicolumn{1}{c}{}\\
    $\sqrt{s}$ [GeV]& $\sigma_{\text{LO}}^{\text{tot}}$ [fb] & $\sigma_{\text{NLO}}^{\text{tot}}$ [fb] & $\sigma_{\text{LO}}^{\text{tot}}$ [fb] & $\sigma_{\text{NLO}}^{\text{tot}}$ [fb] & $\delta_{\text{EW}}$ [\%]&$\sigma^{\text{sig}}$ (LO/NLO)\\
    \hline
    $250$       & $225.59(1)$ &       $ 206.77(1)$      &    $225.60(1)$    & $207.0(1)$& $-8.25$ &$0.4/2.1$\\
    $500$       & $53.74(1)$ &       $62.42(1) $      &    $53.74(3)$    & $62.41(2)$& $+16.14$ &$0.2/0.3$\\
    $1000$  & $12.05(1)$ &     $14.56(1)$   &   $12.0549(6)$ & $14.57(1)$&$+20.84$&$0.5/0.5$ \\
  \end{tabularx}
  \caption[muon collider NLO EW]{Comparison of LO and NLO total inclusive cross sections results of the \texttt{MCSANCee} and \texttt{WHIZARD+RECOLA} setup for the unpolarized process $e^+e^-\rightarrow HZ$}
  \label{eeHZtable}
\end{table}


\section{Derivation of the $HZ$ Sudakov correction factor}
\label{HZapprox}
We derive the analytic form of the NLL EW Sudakov correction factor to
$\mu^+\mu^-\rightarrow HZ$ by applying the general factorization
formalism of \cite{Denner:2000jv,Denner:2001gw}. Due to the
same EW coupling behavior in one-loop processes of
$f\bar{f}\rightarrow HZ$ for fermions $f\neq t$, neglecting all masses
$m_f$, and taking into account the explicit values of the electric
charges $Q_f$, this is in analogy to the results of the generic
flavor- and chirality-dependent formulae to $q\bar{q}\rightarrow HZ$
in \cite{Granata:2017iod}.  For $s\gg M_{W}$ -- using the following
abbreviations for the double and single logarithmic factors,
\begin{align}
  &L(s,M^2_W)=\frac{\alpha}{4\pi}\log^2\frac{s}{M^2_W}
  &l(s,M^2_W)=\frac{\alpha}{4\pi}\log\frac{s}{M^2_W} \qquad ,
\end{align}
we can approximate the leading logarithmic, angular-independent, terms
coming from exchange of soft-collinear gauge bosons between pairs of
external legs, by the term containg double-logarithmic,
single-logarithmic and non-logarithmic contributions,
\begin{align}
  \Lambda^{\kappa}_{l,\lambda}=A^{\kappa}_{\lambda} \; L(s,M_W^2) +
  B^{\kappa}_{\lambda} \; \log\frac{M_Z^2}{M_W^2}l(s,M_W^2)+C_{\lambda}
  \qquad .
  \label{Sud}
\end{align}
Here, $\lambda=T,L$ denote the transverse and longitudinal
polarized $Z$ boson, and $\kappa=L,R$ the muon initial state
chirality, respectively. The constant parameters depend on the quantum
numbers of the external particles and read
\begin{subequations}
  \begin{align}
    A^{\kappa}_{T}&=\;-\frac{1}{2}\left[2C^{ew}_{\mu^{\kappa}}+C^{ew}_{\Phi}+C^{ew}_{ZZ}\right]
    &
    A^{\kappa}_{L}&=\;-\left[C^{ew}_{\mu^{\kappa}}+C^{ew}_{\Phi}\right]
    \\
    B^{\kappa}_{T}&=\;2(I^Z_{\mu_{\kappa}})^2+(I^Z_H)^2
    &
    B^{\kappa}_{L}&=\;2\left[(I^Z_{\mu_{\kappa}})^2+(I^Z_H)^2\right]
    \\
    C_{T}&=\;\delta^{LSC,h}_H
    &
    C_L&=\;\delta^{LSC,h}_H+\delta^{LSC,h}_{\chi} \qquad.
  \end{align}
\end{subequations}
EW Casimir operators $C^{ew}$ as well as
explicit values for $(I^Z)^2$ and $\delta^{LSC,h}$ are extracted from
\cite{Pozzorini:2001rs} and are given by
\begin{subequations}
\begin{align}
	C^{ew}_{\mu^L}&=\;C^{ew}_{\Phi}=\frac{1+2c_w^2}{4s_w^2c_w^2} &
	C^{ew}_{\mu^R}&=\;\frac{1}{c_w^2} &
	C^{ew}_{ZZ}&=\;2\frac{c_w^2}{s_w^2}\\
	(I^Z_{\mu_L})^2&=\;\frac{(c_w^2-s_w^2)^2}{4s_w^2c_w^2} &
	(I^Z_{\mu_R})^2&=\;\frac{s_w^2}{c_w^2} &
	(I^Z_{H})^2&=\;\frac{1}{4s_w^2c_w^2}
\end{align}
\vspace{-15mm}
\begin{align}
  \delta^{LSC,h}_H &=\; \quad\frac{\alpha}{8\pi
    s_w^2} \;\; \, \cdot
  \left[\frac{1}{2c_w^2}  \ln^2 \left(\frac{M_H^2}{M_Z^2}\right)+\ln^2
    \left(\frac{M_H^2}{M_W^2}\right)\right]
  \\
  \delta^{LSC,h}_{\chi} &=\; \frac{\alpha}{16\pi s_w^2c_w^2} \, \cdot \ln^2
  \left(\frac{M_H^2}{M_Z^2}\right) \qquad .
\end{align}
\end{subequations}
In the same context, subleading, angular-dependent, terms
proportional to $l(s,M_W)\log(|t|/s)$ and $l(s,M_W^2)\log(|u|/s)$ due
to $W^{\pm}$ connecting initial- and final-state legs arise. For the
considered Higgsstrahlung process, they take the form\footnote{Note
that the logarithms in the bracket do not become large except for
the extreme forward and backward region, which are anyhow not
experimentally accessible at high-energy muon colliders.}
\begin{align}
  \label{sudangle}
  \Lambda^{\kappa}_{\theta,\lambda}=- \delta_{\kappa L}
  \frac{D_{\lambda}}{I^Z_{\mu_{\kappa}}} \;
  l(s,M_W^2)\left[\log\frac{|t|}{s} +
    \log\frac{|u|}{s}\right]
\end{align}
with constants
\begin{equation}
  D_T =\; - \frac{c_w(1+c_w^2)}{2s_w^2} \qquad
  D_L =\; -\frac{c_w}{s_w} \qquad ,
\end{equation}
using the shortcuts $s_w=\sin \theta_W$ and $c_w=\cos
\theta_W$. Considering the Mandelstam variables $t$ and $u$ in the
high energy limit
\begin{equation}
  t=\;(p_{\mu^+}-p_H)^2\sim -\frac{s}{2}(1-\cos \theta_{H}) \qquad
  u=\;(p_{\mu^+}-p_Z)^2\sim -\frac{s}{2}(1+\cos \theta_{H}) \quad,
\end{equation}
Eq.~(\ref{sudangle}) can be written in terms of the Higgs polar angle
$\theta_{H}$. Single-logarithmic terms $\Lambda^{\kappa}_{s,\lambda}$ originating from virtual
soft/collinear gauge bosons emitted from single external legs
(wave-function renormalization diagrams) and $\Lambda^{\kappa}_{\text{PR},\lambda}$ from renormalization of coupling
parameters can be expressed as
\begin{equation}
  \Lambda^{\kappa}_{s,\lambda} =\; E^{\kappa}_{s,\lambda} \;
  l(s,M_W^2) + \frac{\alpha}{4\pi} F^{\kappa}_{s,\lambda}
  \qquad
  \Lambda^{\kappa}_{\text{PR},\lambda} =\;
  E^{\kappa}_{\text{PR},\lambda} \; l(s,M_W^2) +
  \frac{\alpha}{4\pi}F^{\kappa}_{\text{PR},\lambda} \quad ,
  \label{singlePR}
\end{equation}
respectively. We give here for completeness the explicit expression
for the quantities $E^{\kappa}_{s,\lambda}$, $F^{\kappa}_{s,\lambda}$,
$E^{\kappa}_{\text{PR},\lambda}$ and $F^{\kappa}_{\text{PR},\lambda}$:
\begin{subequations}
\begin{align}
  &E^{\kappa}_{s,T} = 3C^{ew}_{\mu^{\kappa}}+2C^{ew}_{\Phi}+\frac{1}{2}b^{ew}_{ZZ}
  - \frac{3}{4s_w^2}\frac{m_t^2}{M_W^2}
  \\
  &E^{\kappa}_{s,L} = 3C^{ew}_{\mu^{\kappa}}+4C^{ew}_{\Phi}-\frac{3}{2s_w^2}
  \frac{m_t^2}{M_W^2}
  \\
  &F^{\kappa}_{s,T} = \left(\frac{3}{4s_w^2}\frac{m_t^2}{M_W^2}+T_{ZZ}\right)\log
  \frac{m_t^2}{M_W^2}+\left(\frac{M_Z^2}{24s_w^2M_W^2}-2C^{ew}_{\Phi}\right)\log
  \frac{M_H^2}{M_W^2}
  \\
  &F^{\kappa}_{s,L} = \frac{3}{2s_w^2}\frac{m_t^2}{M_W^2}\log
  \frac{m_t^2}{M_W^2}+\left(\frac{M_Z^2}{8s_w^2M_W^2}-2C^{ew}_{\Phi}\right)\log
  \frac{M_H^2}{M_W^2}
  \\
  &E^{\kappa}_{\text{PR},T} =
  -b^{ew}_{WW}+\rho_{\mu_{\kappa}}\frac{s_w}{c_w}b^{ew}_{AZ}+2C^{ew}_{\Phi}
  - \frac{1}{2}b^{ew}_{ZZ}-\frac{3}{4s_w^2}\frac{m_t^2}{M_W^2}
  \\
  &E^{\kappa}_{\text{PR},L}=-b^{ew}_{WW}+\rho_{\mu_{\kappa}}\frac{s_w}{c_w}b^{ew}_{AZ}
  \\
  \begin{split}
    &F^{\kappa}_{\text{PR},T}=\frac{5}{6}\left(\frac{1}{s_w^2} +
    \frac{\rho_{\mu_{\kappa}}}{c_w^2}-\frac{M_Z^2}{2s_w^2M_W^2}\right)\log
    \frac{M_H^2}{M_W^2}
    \\
    &\qquad\qquad-\left[\frac{9+6s_w^2-32s^4_w}{18s_w^2}\left(\frac{1}{s_w^2}
      + \frac{\rho_{\mu_{\kappa}}}{c_w^2}\right)+T_{ZZ} -
      \frac{3}{4s_w^2}\frac{m_t^2}{M_W^2}\right]\log
    \frac{m_t^2}{M_W^2}
  \end{split}
  \\
  &F^{\kappa}_{\text{PR},L}=\left(\frac{1}{s_w^2} +
  \frac{\rho_{\mu_{\kappa}}}{c_w^2}\right)\left[\frac{5}{6}\log
    \frac{M_H^2}{M_W^2} -\frac{9+6s_w^2-32s_w^4}{18s_w^2}\log
    \frac{m_t^2}{M_W^2} \right]
\end{align}
\end{subequations}
The explicit values of the $\beta$-function coefficients
$b^{ew}_{ZZ}$, $b^{ew}_{AZ}$ and $b^{ew}_{WW}$ and the numerical
coefficient $T_{ZZ}$ used here are (cf.~\cite{Pozzorini:2001rs}):
\begin{subequations}
  \begin{align}
    &b^{ew}_{ZZ}=\frac{19-38s_w^2-22s_w^4}{6s_w^2c_w^2} &
    b^{ew}_{AZ}=-\frac{19+22s_w^2}{6s_wc_w}&
    &b^{ew}_{WW}=\frac{19}{6s_w^2}\\
    &T_{ZZ}=\frac{9-24s_w^2+32s_w^4}{36s_w^2c_w^2} \qquad.
  \end{align}
\end{subequations}
The constants $E^{\kappa}_{\text{PR},\lambda}$ and
$F^{\kappa}_{\text{PR},\lambda}$ depend on the parameter
$\rho_{\mu_{\kappa}}$ which is defined by the EW quantum numbers of
the muon as \cite{Granata:2017iod}
\begin{align}
  \rho_{\mu_{\kappa}}=\frac{Q_{\mu}-T^3_{\mu_{\kappa}}}{T^3_{\mu_{\kappa}}-Q_{\mu}s_w^2}
  \qquad.
\end{align}
Summing the contributions from Eqs.~(\ref{Sud}), (\ref{sudangle}) and
(\ref{singlePR}), the overall Sudakov factor as a function of the
muon chirality $\kappa$ and $Z$ boson polarization $\lambda$ can be
formulated as
\begin{equation}
  \label{sudoverall}
  \Lambda^{\kappa}_{\lambda} = \Lambda^{\kappa}_{l,\lambda} +
  \Lambda^{\kappa}_{\theta,\lambda} + \Lambda^{\kappa}_{s,\lambda} +
  \Lambda^{\kappa}_{\text{PR},\lambda} \qquad .
\end{equation}
This factor can now be used for an approximation of the relative NLO
correction to the unpolarized process for which the photon radiation
effects are subtracted, i.~e. emulating the virtual EW effects in the
high-energy limit. For the definition of this correction factor we apply
two different approximations. The first one estimates
\begin{equation}
  \label{longitudinalestimation}
  \Lambda^{\kappa}_{\lambda}\,\mathcal{M}^{\mu^+_{\kappa}\mu^-_{\kappa}\rightarrow
    HZ_{\lambda}}_0 \quad \xrightarrow{s\gg M_W^2} \quad
  \delta_{\lambda L}\Lambda^{\kappa}_{\lambda} \,
  \mathcal{M}^{\mu^+_{\kappa}\mu^-_{\kappa}\rightarrow HZ_{\lambda}}_0
\end{equation}
since the Born amplitudes for transverse
polarized $Z$ bosons are suppressed by $M_Z^2/s$
\cite{Chanowitz:1985hj,Bohm:2001yx}.

For the second approximation, we assume amplitudes with the helicity
configurations $(+,+)$ and $(-,-)$ of the muons to vanish due to the
ultra-relativistic initial state momenta in the process. We thus
arrive at a factor of the form
\begin{align}
  \label{unpolsud}
  \Lambda_{\text{est}}^{\text{unpol}} =
  \frac{\sum_{\kappa}\Lambda^{\kappa}_{L}
    \left|\mathcal{M}^{\mu^+_{\kappa}\mu^-_{\kappa}\rightarrow
      HZ_{L}}_0\right|^2}{4\left|\mathcal{M}^{\mu^+\mu^-\rightarrow
      HZ_L}_0\right|^2}
\end{align}
which approximates the non-integrated virtual corrections for
unpolarized beams in the high-energy limit, when multiplied with
unpolarized Born squared amplitudes $\left|\mathcal{M}^{\mu^+\mu^-\rightarrow
HZ_L}_0\right|^2$. Using the fact that the amplitudes depend on the
muon chiralities only through the EW couplings, the angular dependency
of squared Born amplitudes exclusive in the muon chiralities normalized
to the unpolarized ones, drops out. From this, we find
\begin{align}
  \frac{\left|\mathcal{M}^{\mu^+_{\kappa}\mu^-_{\kappa}\rightarrow
      HZ_{L}}_0\right|^2}{\left|\mathcal{M}^{\mu^+\mu^-\rightarrow
      HZ^L}_0\right|^2}=\frac{\left(d\sigma^{\mu^+_{\kappa}\mu^-_{\kappa}\rightarrow
      HZ_{L}}_{B}/d\Omega\right)}{\left(d\sigma^{\mu^+\mu^-\rightarrow
      HZ_{L}}_{B}/d\Omega\right)}=\frac{\sigma^{\mu^+_{\kappa}\mu^-_{\kappa}\rightarrow
      HZ_{L}}_{B}}{\sigma^{\mu^+\mu^-\rightarrow
      HZ_{L}}_{B}}= \frac{\sigma^{\mu^+_{\kappa}\mu^-_{\kappa}\rightarrow
      HZ}_{B}}{\sigma^{\mu^+\mu^-\rightarrow
      HZ}_{B}}
  \quad .
\end{align}
Hence, in our approach Eq.~(\ref{unpolsud}) can be rewritten as
\begin{align}
  \Lambda_{\text{est}}^{\text{unpol}} = \frac{1}{4}
  \frac{\sum_{\kappa} \, \Lambda^{\kappa}_{L} \,
    \sigma^{\mu^+_{\kappa}\mu^-_{\kappa}\rightarrow
      HZ}_{B}}{\sigma^{\mu^+\mu^-\rightarrow
      HZ}_{B}} \qquad.
  \label{finaldeltaunpol}
\end{align}
For the evaluation of this approximative correction factor, integrated
polarized and unpolarized Born cross sections $\sigma_B$
obtained with the \texttt{WHIZARD+RECOLA} framework are used. The
analytical Sudakov factors from Eqs.~(\ref{sudoverall}) and the factor
of Eq.~(\ref{finaldeltaunpol}) for the (central) scattering angle
$\theta_{H}=90^{\circ}$ as well as $\Lambda^{\text{unpol}}_{\text{est,c}}$,
i.~e. Eq.~(\ref{finaldeltaunpol}) with the angular-dependent terms
$\Lambda^{\kappa}_{\theta,L}$ dropped, are depicted in
Fig.~\ref{sudpic} as a function of the center-of-mass energy in the
main text. There, in Sec.~\ref{scans} , also the physics implications
of the Sudakov factors and their comparison with the complete
fixed-order NLO EW cross sections are discussed.

\bibliographystyle{JHEP}
\bibliography{mu}

\providecommand{\href}[2]{#2}\begingroup\raggedright\begin{thebibliography}{10}

\bibitem{Abada:2019lih}
{\scshape FCC} collaboration, \emph{{FCC Physics Opportunities}: {Future
  Circular Collider Conceptual Design Report Volume 1}},
  \href{https://doi.org/10.1140/epjc/s10052-019-6904-3}{\emph{Eur. Phys. J. C}
  {\bfseries 79} (2019) 474}.

\bibitem{Benedikt:2018csr}
{\scshape FCC} collaboration, \emph{{FCC-hh: The Hadron Collider}: {Future
  Circular Collider Conceptual Design Report Volume 3}},
  \href{https://doi.org/10.1140/epjst/e2019-900087-0}{\emph{Eur. Phys. J. ST}
  {\bfseries 228} (2019) 755}.

\bibitem{Baer:2013cma}
H.~Baer et~al., \emph{{The International Linear Collider Technical Design
  Report - Volume 2: Physics}},
  \href{https://arxiv.org/abs/1306.6352}{{\ttfamily 1306.6352}}.

\bibitem{Behnke:2013lya}
H.~Abramowicz et~al., \emph{{The International Linear Collider Technical Design
  Report - Volume 4: Detectors}},
  \href{https://arxiv.org/abs/1306.6329}{{\ttfamily 1306.6329}}.

\bibitem{CLIC:2016zwp}
{\scshape CLIC, CLICdp} collaboration, \emph{{Updated baseline for a staged
  Compact Linear Collider}},
  \href{https://arxiv.org/abs/1608.07537}{{\ttfamily 1608.07537}}.

\bibitem{Aicheler:2012bya}
M.~Aicheler et~al., \emph{{A Multi-TeV Linear Collider Based on CLIC
  Technology}: {CLIC Conceptual Design Report}},
  \href{https://arxiv.org/abs/CERN-2012-007}{{\ttfamily CERN-2012-007}}.

\bibitem{Delahaye:2019omf}
J.P.~Delahaye, M.~Diemoz, K.~Long, B.~Mansouli\'e, N.~Pastrone, L.~Rivkin
  et~al., \emph{{Muon Colliders}},
  \href{https://arxiv.org/abs/1901.06150}{{\ttfamily 1901.06150}}.

\bibitem{Bartosik:2020xwr}
N.~Bartosik et~al., \emph{{Detector and Physics Performance at a Muon
  Collider}},
  \href{https://doi.org/10.1088/1748-0221/15/05/P05001}{\emph{JINST} {\bfseries
  15} (2020) P05001} [\href{https://arxiv.org/abs/2001.04431}{{\ttfamily
  2001.04431}}].

\bibitem{Schulte:2020xvf}
D.~Schulte, J.-P.~Delahaye, M.~Diemoz, K.~Long, B.~Mansouli\'e, N.~Pastrone
  et~al., \emph{{Prospects on Muon Colliders}},
  \href{https://doi.org/10.22323/1.390.0703}{\emph{PoS} {\bfseries ICHEP2020}
  (2021) 703}.

\bibitem{Long:2021upy}
K.R.~Long, D.~Lucchesi, M.A.~Palmer, N.~Pastrone, D.~Schulte and V.~Shiltsev,
  \emph{{Muon colliders to expand frontiers of particle physics}},
  \href{https://arxiv.org/abs/FERMILAB-PUB-21-040-AD}{{\ttfamily
  FERMILAB-PUB-21-040-AD}}.

\bibitem{Aime:2022flm}
C.~Aime et~al., \emph{{Muon Collider Physics Summary}},
  \href{https://arxiv.org/abs/2203.07256}{{\ttfamily 2203.07256}}.

\bibitem{MuonCollider:2022xlm}
{\scshape Muon Collider} collaboration, \emph{{The physics case of a 3 TeV muon
  collider stage}},  \href{https://arxiv.org/abs/2203.07261}{{\ttfamily
  2203.07261}}.

\bibitem{CLIC:2018fvx}
{\scshape CLIC} collaboration, \emph{{The CLIC Potential for New Physics}},
  \href{https://arxiv.org/abs/1812.02093}{{\ttfamily 1812.02093}}.

\bibitem{Costantini:2020stv}
A.~Costantini, F.~De~Lillo, F.~Maltoni, L.~Mantani, O.~Mattelaer, R.~Ruiz
  et~al., \emph{{Vector boson fusion at multi-TeV muon colliders}},
  \href{https://doi.org/10.1007/JHEP09(2020)080}{\emph{JHEP} {\bfseries 09}
  (2020) 080} [\href{https://arxiv.org/abs/2005.10289}{{\ttfamily
  2005.10289}}].

\bibitem{Han:2020pif}
T.~Han, D.~Liu, I.~Low and X.~Wang, \emph{{Electroweak couplings of the Higgs
  boson at a multi-TeV muon collider}},
  \href{https://doi.org/10.1103/PhysRevD.103.013002}{\emph{Phys. Rev. D}
  {\bfseries 103} (2021) 013002}
  [\href{https://arxiv.org/abs/2008.12204}{{\ttfamily 2008.12204}}].

\bibitem{Chiesa:2020awd}
M.~Chiesa, F.~Maltoni, L.~Mantani, B.~Mele, F.~Piccinini and X.~Zhao,
  \emph{{Measuring the quartic Higgs self-coupling at a multi-TeV muon
  collider}}, \href{https://doi.org/10.1007/JHEP09(2020)098}{\emph{JHEP}
  {\bfseries 09} (2020) 098}
  [\href{https://arxiv.org/abs/2003.13628}{{\ttfamily 2003.13628}}].

\bibitem{Dermisek:2021mhi}
R.~Dermisek, K.~Hermanek and N.~McGinnis, \emph{{Di-Higgs and tri-Higgs boson
  signals of muon g-2 at a muon collider}},
  \href{https://doi.org/10.1103/PhysRevD.104.L091301}{\emph{Phys. Rev. D}
  {\bfseries 104} (2021) L091301}
  [\href{https://arxiv.org/abs/2108.10950}{{\ttfamily 2108.10950}}].

\bibitem{Han:2021lnp}
T.~Han, W.~Kilian, N.~Kreher, Y.~Ma, J.~Reuter, T.~Striegl et~al.,
  \emph{{Precision test of the muon-Higgs coupling at a high-energy muon
  collider}}, \href{https://doi.org/10.1007/JHEP12(2021)162}{\emph{JHEP}
  {\bfseries 12} (2021) 162}
  [\href{https://arxiv.org/abs/2108.05362}{{\ttfamily 2108.05362}}].

\bibitem{Kilian:2007gr}
W.~Kilian, T.~Ohl and J.~Reuter, \emph{{WHIZARD: Simulating Multi-Particle
  Processes at LHC and ILC}},
  \href{https://doi.org/10.1140/epjc/s10052-011-1742-y}{\emph{Eur. Phys. J. C}
  {\bfseries 71} (2011) 1742}
  [\href{https://arxiv.org/abs/0708.4233}{{\ttfamily 0708.4233}}].

\bibitem{Moretti:2001zz}
M.~Moretti, T.~Ohl and J.~Reuter, \emph{{O'Mega: An Optimizing matrix element
  generator}},  \href{https://arxiv.org/abs/hep-ph/0102195}{{\ttfamily
  hep-ph/0102195}}.

\bibitem{WhizardNLO}
S.~Bra\ss, P.~Bredt, W.~Kilian, J.~Reuter, V.~Rothe and P.~Stienemeier,
  \emph{{Automation of NLO SM processes in WHIZARD for hadron and lepton
  collisions, in preparation}},  2022.

\bibitem{Actis:2016mpe}
S.~Actis, A.~Denner, L.~Hofer, J.-N.~Lang, A.~Scharf and S.~Uccirati,
  \emph{{RECOLA: REcursive Computation of One-Loop Amplitudes}},
  \href{https://doi.org/10.1016/j.cpc.2017.01.004}{\emph{Comput. Phys. Commun.}
  {\bfseries 214} (2017) 140}
  [\href{https://arxiv.org/abs/1605.01090}{{\ttfamily 1605.01090}}].

\bibitem{Buccioni:2019sur}
F.~Buccioni, J.-N.~Lang, J.M.~Lindert, P.~Maierh\"ofer, S.~Pozzorini, H.~Zhang
  et~al., \emph{{OpenLoops 2}},
  \href{https://doi.org/10.1140/epjc/s10052-019-7306-2}{\emph{Eur. Phys. J. C}
  {\bfseries 79} (2019) 866}
  [\href{https://arxiv.org/abs/1907.13071}{{\ttfamily 1907.13071}}].

\bibitem{Frixione:1995ms}
S.~Frixione, Z.~Kunszt and A.~Signer, \emph{{Three jet cross-sections to
  next-to-leading order}},
  \href{https://doi.org/10.1016/0550-3213(96)00110-1}{\emph{Nucl. Phys. B}
  {\bfseries 467} (1996) 399}
  [\href{https://arxiv.org/abs/hep-ph/9512328}{{\ttfamily hep-ph/9512328}}].

\bibitem{Frixione:1997np}
S.~Frixione, \emph{{A General approach to jet cross-sections in QCD}},
  \href{https://doi.org/10.1016/S0550-3213(97)00574-9}{\emph{Nucl. Phys. B}
  {\bfseries 507} (1997) 295}
  [\href{https://arxiv.org/abs/hep-ph/9706545}{{\ttfamily hep-ph/9706545}}].

\bibitem{Kilian:2012pz}
W.~Kilian, T.~Ohl, J.~Reuter and C.~Speckner, \emph{{QCD in the Color-Flow
  Representation}}, \href{https://doi.org/10.1007/JHEP10(2012)022}{\emph{JHEP}
  {\bfseries 10} (2012) 022} [\href{https://arxiv.org/abs/1206.3700}{{\ttfamily
  1206.3700}}].

\bibitem{ChokoufeNejad:2016qux}
B.~Chokouf\'e~Nejad, W.~Kilian, J.M.~Lindert, S.~Pozzorini, J.~Reuter and
  C.~Weiss, \emph{{NLO QCD predictions for off-shell $ t\overline{t} $ and $
  t\overline{t}H $ production and decay at a linear collider}},
  \href{https://doi.org/10.1007/JHEP12(2016)075}{\emph{JHEP} {\bfseries 12}
  (2016) 075} [\href{https://arxiv.org/abs/1609.03390}{{\ttfamily
  1609.03390}}].

\bibitem{Kilian:2006cj}
W.~Kilian, J.~Reuter and T.~Robens, \emph{{NLO Event Generation for Chargino
  Production at the ILC}},
  \href{https://doi.org/10.1140/epjc/s10052-006-0048-y}{\emph{Eur. Phys. J. C}
  {\bfseries 48} (2006) 389}
  [\href{https://arxiv.org/abs/hep-ph/0607127}{{\ttfamily hep-ph/0607127}}].

\bibitem{Robens:2008sa}
T.~Robens, J.~Kalinowski, K.~Rolbiecki, W.~Kilian and J.~Reuter, \emph{{(N)LO
  Simulation of Chargino Production and Decay}}, {\emph{Acta Phys. Polon. B}
  {\bfseries 39} (2008) 1705}
  [\href{https://arxiv.org/abs/0803.4161}{{\ttfamily 0803.4161}}].

\bibitem{Binoth:2009rv}
T.~Binoth, N.~Greiner, A.~Guffanti, J.~Reuter, J.P.~Guillet and T.~Reiter,
  \emph{{Next-to-leading order QCD corrections to pp --\ensuremath{>} b anti-b
  b anti-b + X at the LHC: the quark induced case}},
  \href{https://doi.org/10.1016/j.physletb.2010.02.010}{\emph{Phys. Lett. B}
  {\bfseries 685} (2010) 293}
  [\href{https://arxiv.org/abs/0910.4379}{{\ttfamily 0910.4379}}].

\bibitem{Greiner:2011mp}
N.~Greiner, A.~Guffanti, T.~Reiter and J.~Reuter, \emph{{NLO QCD corrections to
  the production of two bottom-antibottom pairs at the LHC}},
  \href{https://doi.org/10.1103/PhysRevLett.107.102002}{\emph{Phys. Rev. Lett.}
  {\bfseries 107} (2011) 102002}
  [\href{https://arxiv.org/abs/1105.3624}{{\ttfamily 1105.3624}}].

\bibitem{Bach:2017ggt}
F.~Bach, B.C.~Nejad, A.~Hoang, W.~Kilian, J.~Reuter, M.~Stahlhofen et~al.,
  \emph{{Fully-differential Top-Pair Production at a Lepton Collider: From
  Threshold to Continuum}},
  \href{https://doi.org/10.1007/JHEP03(2018)184}{\emph{JHEP} {\bfseries 03}
  (2018) 184} [\href{https://arxiv.org/abs/1712.02220}{{\ttfamily
  1712.02220}}].

\bibitem{Dittmaier:2015bfe}
S.~Dittmaier and C.~Schwan, \emph{{Non-factorizable photonic corrections to
  resonant production and decay of many unstable particles}},
  \href{https://doi.org/10.1140/epjc/s10052-016-3968-1}{\emph{Eur. Phys. J. C}
  {\bfseries 76} (2016) 144}
  [\href{https://arxiv.org/abs/1511.01698}{{\ttfamily 1511.01698}}].

\bibitem{Denner:2000bj}
A.~Denner, S.~Dittmaier, M.~Roth and D.~Wackeroth, \emph{{Electroweak radiative
  corrections to e+ e- ---\ensuremath{>} W W ---\ensuremath{>} 4 fermions in
  double pole approximation: The RACOONWW approach}},
  \href{https://doi.org/10.1016/S0550-3213(00)00511-3}{\emph{Nucl. Phys. B}
  {\bfseries 587} (2000) 67}
  [\href{https://arxiv.org/abs/hep-ph/0006307}{{\ttfamily hep-ph/0006307}}].

\bibitem{Ohl:1998jn}
T.~Ohl, \emph{{Vegas revisited: Adaptive Monte Carlo integration beyond
  factorization}},
  \href{https://doi.org/10.1016/S0010-4655(99)00209-X}{\emph{Comput. Phys.
  Commun.} {\bfseries 120} (1999) 13}
  [\href{https://arxiv.org/abs/hep-ph/9806432}{{\ttfamily hep-ph/9806432}}].

\bibitem{Brass:2018xbv}
S.~Brass, W.~Kilian and J.~Reuter, \emph{{Parallel Adaptive Monte Carlo
  Integration with the Event Generator WHIZARD}},
  \href{https://doi.org/10.1140/epjc/s10052-019-6840-2}{\emph{Eur. Phys. J. C}
  {\bfseries 79} (2019) 344}
  [\href{https://arxiv.org/abs/1811.09711}{{\ttfamily 1811.09711}}].

\bibitem{Sadykov:2020dgm}
R.R.~Sadykov, A.B.~Arbuzov, S.G.~Bondarenko, Y.V.~Dydyshka, L.V.~Kalinovskaya,
  I.I.~Novikov et~al., \emph{{MCSANCee generator with one-loop electroweak
  corrections for processes with polarized e+e- beams}},
  \href{https://doi.org/10.1088/1742-6596/1525/1/012012}{\emph{J. Phys. Conf.
  Ser.} {\bfseries 1525} (2020) 012012}.

\bibitem{Nogueira:1992en}
P.~Nogueira and J.C.~Romao, \emph{{Initial-final state interference in e+ e-
  ---\ensuremath{>} H mu+ mu-}},
  \href{https://doi.org/10.1007/BF01558407}{\emph{Z. Phys. C} {\bfseries 60}
  (1993) 757}.

\bibitem{Kniehl:1993ay}
B.A.~Kniehl, \emph{{Higgs phenomenology at one loop in the standard model}},
  \href{https://doi.org/10.1016/0370-1573(94)90037-X}{\emph{Phys. Rept.}
  {\bfseries 240} (1994) 211}.

\bibitem{Belanger:2002ik}
G.~Belanger, F.~Boudjema, J.~Fujimoto, T.~Ishikawa, T.~Kaneko, K.~Kato et~al.,
  \emph{{Full one loop electroweak radiative corrections to single Higgs
  production in e+ e-}},
  \href{https://doi.org/10.1016/S0370-2693(03)00339-3}{\emph{Phys. Lett. B}
  {\bfseries 559} (2003) 252}
  [\href{https://arxiv.org/abs/hep-ph/0212261}{{\ttfamily hep-ph/0212261}}].

\bibitem{Song:2021vru}
Q.~Song and A.~Freitas, \emph{{On the evaluation of two-loop electroweak box
  diagrams for $e^+e^- \to HZ$ production}},
  \href{https://doi.org/10.1007/JHEP04(2021)179}{\emph{JHEP} {\bfseries 04}
  (2021) 179} [\href{https://arxiv.org/abs/2101.00308}{{\ttfamily
  2101.00308}}].

\bibitem{Kuhn:1999nn}
J.H.~Kuhn, A.A.~Penin and V.A.~Smirnov, \emph{{Summing up subleading Sudakov
  logarithms}}, \href{https://doi.org/10.1007/s100520000462}{\emph{Eur. Phys.
  J. C} {\bfseries 17} (2000) 97}
  [\href{https://arxiv.org/abs/hep-ph/9912503}{{\ttfamily hep-ph/9912503}}].

\bibitem{Denner:2000jv}
A.~Denner and S.~Pozzorini, \emph{{One loop leading logarithms in electroweak
  radiative corrections. 1. Results}},
  \href{https://doi.org/10.1007/s100520100551}{\emph{Eur. Phys. J. C}
  {\bfseries 18} (2001) 461}
  [\href{https://arxiv.org/abs/hep-ph/0010201}{{\ttfamily hep-ph/0010201}}].

\bibitem{Denner:2001gw}
A.~Denner and S.~Pozzorini, \emph{{One loop leading logarithms in electroweak
  radiative corrections. 2. Factorization of collinear singularities}},
  \href{https://doi.org/10.1007/s100520100721}{\emph{Eur. Phys. J. C}
  {\bfseries 21} (2001) 63}
  [\href{https://arxiv.org/abs/hep-ph/0104127}{{\ttfamily hep-ph/0104127}}].

\bibitem{Bell:2010gi}
G.~Bell, J.H.~Kuhn and J.~Rittinger, \emph{{Electroweak Sudakov Logarithms and
  Real Gauge-Boson Radiation in the TeV Region}},
  \href{https://doi.org/10.1140/epjc/s10052-010-1489-x}{\emph{Eur. Phys. J. C}
  {\bfseries 70} (2010) 659} [\href{https://arxiv.org/abs/1004.4117}{{\ttfamily
  1004.4117}}].

\bibitem{Granata:2017iod}
F.~Granata, J.M.~Lindert, C.~Oleari and S.~Pozzorini, \emph{{NLO QCD+EW
  predictions for HV and HV +jet production including parton-shower effects}},
  \href{https://doi.org/10.1007/JHEP09(2017)012}{\emph{JHEP} {\bfseries 09}
  (2017) 012} [\href{https://arxiv.org/abs/1706.03522}{{\ttfamily
  1706.03522}}].

\bibitem{Chanowitz:1985hj}
M.S.~Chanowitz and M.K.~Gaillard, \emph{{The TeV Physics of Strongly
  Interacting W's and Z's}},
  \href{https://doi.org/10.1016/0550-3213(85)90580-2}{\emph{Nucl. Phys. B}
  {\bfseries 261} (1985) 379}.

\bibitem{Bohm:2001yx}
M.~Bohm, A.~Denner and H.~Joos, \emph{{Gauge theories of the strong and
  electroweak interaction}}, Teubner Verlag; 3rd rev. ed. 2001 Edition (2001),
  \href{https://doi.org/10.1007/978-3-322-80160-9}{10.1007/978-3-322-80160-9}.

\bibitem{Denner:1988tv}
A.~Denner and T.~Sack, \emph{{ELECTROWEAK RADIATIVE CORRECTIONS TO e+ e-
  ---\ensuremath{>} Z0 Z0}},
  \href{https://doi.org/10.1016/0550-3213(88)90691-8}{\emph{Nucl. Phys. B}
  {\bfseries 306} (1988) 221}.

\bibitem{Cacciari:1992pz}
M.~Cacciari, A.~Deandrea, G.~Montagna and O.~Nicrosini, \emph{{QED structure
  functions: A Systematic approach}},
  \href{https://doi.org/10.1209/0295-5075/17/2/007}{\emph{EPL} {\bfseries 17}
  (1992) 123}.

\bibitem{Skrzypek:1990qs}
M.~Skrzypek and S.~Jadach, \emph{{Exact and approximate solutions for the
  electron nonsinglet structure function in QED}},
  \href{https://doi.org/10.1007/BF01483573}{\emph{Z. Phys. C} {\bfseries 49}
  (1991) 577}.

\bibitem{Skrzypek:1992vk}
M.~Skrzypek, \emph{{Leading logarithmic calculations of QED corrections at
  LEP}}, {\emph{Acta Phys. Polon. B} {\bfseries 23} (1992) 135}.

\bibitem{Frixione:2019lga}
S.~Frixione, \emph{{Initial conditions for electron and photon structure and
  fragmentation functions}},
  \href{https://doi.org/10.1007/JHEP11(2019)158}{\emph{JHEP} {\bfseries 11}
  (2019) 158} [\href{https://arxiv.org/abs/1909.03886}{{\ttfamily
  1909.03886}}].

\bibitem{Bertone:2019hks}
V.~Bertone, M.~Cacciari, S.~Frixione and G.~Stagnitto, \emph{{The partonic
  structure of the electron at the next-to-leading logarithmic accuracy in
  QED}}, \href{https://doi.org/10.1007/JHEP03(2020)135}{\emph{JHEP} {\bfseries
  03} (2020) 135} [\href{https://arxiv.org/abs/1911.12040}{{\ttfamily
  1911.12040}}].

\bibitem{Bertone:2022ktl}
V.~Bertone, M.~Cacciari, S.~Frixione, G.~Stagnitto, M.~Zaro and X.~Zhao,
  \emph{{Improving methods and predictions at high-energy $e^+e^-$ colliders
  within collinear factorisation}},
  \href{https://arxiv.org/abs/2207.03265}{{\ttfamily 2207.03265}}.

\bibitem{Han:2020uid}
T.~Han, Y.~Ma and K.~Xie, \emph{{High Energy Leptonic Collisions and
  Electroweak Parton Distribution Functions}},
  \href{https://arxiv.org/abs/2007.14300}{{\ttfamily 2007.14300}}.

\bibitem{Beenakker:1993tt}
W.~Beenakker, A.~Denner, S.~Dittmaier, R.~Mertig and T.~Sack,
  \emph{{High-energy approximation for on-shell W pair production}},
  \href{https://doi.org/10.1016/0550-3213(93)90434-Q}{\emph{Nucl. Phys. B}
  {\bfseries 410} (1993) 245}.

\bibitem{Beenakker:1994vn}
W.~Beenakker and A.~Denner, \emph{{Standard model predictions for $W$ pair
  production in electron - positron collisions}},
  \href{https://doi.org/10.1142/S0217751X94001965}{\emph{Int. J. Mod. Phys. A}
  {\bfseries 9} (1994) 4837}.

\bibitem{Berggren:2022}
{Berggren, M.}, \emph{private communication},  2022.

\bibitem{NLO_multEW}
P.~Bredt, W.~Kilian, K.~M\k{e}ka{\l}a, J.~Reuter and A.~\.Zarnecki,
  \emph{{Precision study of multi-boson processes at a muon collider, in
  preparation}},  2022.

\bibitem{Pozzorini:2001rs}
S.~Pozzorini, \emph{{Electroweak radiative corrections at high-energies}},
  other thesis, {University of Z\"urich}, 2001,
  [\href{https://arxiv.org/abs/hep-ph/0201077}{{\ttfamily hep-ph/0201077}}].

\end{thebibliography}\endgroup

\end{document}